\documentclass[preprint2]{aastex62}
\usepackage{longtable}
\usepackage{multirow}
\usepackage{graphicx}           
\usepackage{amsmath,amssymb}
\usepackage[T1]{fontenc}

\graphicspath{{./}{figures/}}

\submitjournal{ApJ}

\shorttitle{iMOGABA: OJ~287}
\shortauthors{Lee et al.}

\begin{document}
\title{Interferometric Monitoring of Gamma-ray Bright AGNs: OJ\,287}
\correspondingauthor{Jee Won Lee}
\email{jwlee78@kasi.re.kr}

\author[0000-0002-0786-7307]{Jee Won Lee}
\affil{Korea Astronomy and Space Science Institute, 
776 Daedeok-daero, Yuseong-gu, Daejeon 34055,
Republic of Korea; Correspondence to Jee Won Lee at jwlee78@kasi.re.kr}
\author{Sang-Sung Lee}
\affil{Korea Astronomy and Space Science Institute, 
776 Daedeok-daero, Yuseong-gu, Daejeon 34055,
Republic of Korea; Correspondence to Jee Won Lee at jwlee78@kasi.re.kr}
\affil{University of Science and Technology, 217 Gajeong-ro, Yuseong-gu, Daejeon 34113, Korea}
\author{Juan-Carlos Algaba}
\affil{Department of Physics, Faculty of Science, University of Malaya, 50603 Kuala Lumpur, Malaysia}
\author{Jeffrey Hodgson}
\affil{Korea Astronomy and Space Science Institute, 
776 Daedeok-daero, Yuseong-gu, Daejeon 34055,
Republic of Korea; Correspondence to Jee Won Lee at jwlee78@kasi.re.kr}
\affil{Department of Physics and Astronomy, Sejong University, 209 Neungdong-ro, Gwangjin-gu, Seoul, South Korea}
\author{Jae-Young Kim}
\affil{Max-Planck-Institut f$\"{u}$r Radioastronomie, Auf dem H$\"{u}$gel 69, D-53121 Bonn, Germany}
\author{Jongho Park}
\affil{Institute of Astronomy and Astrophysics, Academia Sinica, P.O. Box 23-141, Taipei 10617, Taiwan}
\author{Motoki Kino}
\affil{National Astronomical Observatory of Japan, 2-21-1 Osawa, Mitaka, Tokyo 181-8588, Japan}
\affil{Kogakuin University of Technology \& Engineering, Academic Support Center, 2665-1 Nakano, Hachioji,  Tokyo 192-0015, Japan}
\author{Dae-Won Kim}
\affil{Department of Physics and Astronomy, Seoul National University, Gwanak-gu, Seoul 08826, Korea}
\author{Sincheol Kang}
\affil{Korea Astronomy and Space Science Institute, 
776 Daedeok-daero, Yuseong-gu, Daejeon 34055,
Republic of Korea; Correspondence to Jee Won Lee at jwlee78@kasi.re.kr}
\affil{University of Science and Technology, 217 Gajeong-ro, Yuseong-gu, Daejeon 34113, Korea}
\author{Sungmin Yoo}
\affil{Department of astronomy and space science, Chungbuk National University, Cheongju, 28644, Republic of Korea}
\author{Sang Hyun Kim}
\affil{Korea Astronomy and Space Science Institute, 
776 Daedeok-daero, Yuseong-gu, Daejeon 34055,
Republic of Korea; Correspondence to Jee Won Lee at jwlee78@kasi.re.kr}
\affil{University of Science and Technology, 217 Gajeong-ro, Yuseong-gu, Daejeon 34113, Korea}
\author{Mark Gurwell}
\affil{Center for Astrophysics | Harvard $\&$ Smithsonian, 60 Garden Street, Cambridge, MA 02138 USA}

\begin{abstract}
We present the results of simultaneous multi-frequency imaging observations at 22, 43, 86, and 129\,GHz of OJ\,287. We used the Korean VLBI Network as part of the Interferometric MOnitoring of GAmma-ray Bright active galactic nuclei (iMOGABA).
The iMOGABA observations were performed during 31 epochs from 2013 January 16 to 2016 December 28. We also used 15\,GHz OVRO and 225\,GHz SMA flux density data. We analyzed four flux enhancements in the light curves. 
The estimated time scales of three flux enhancements were similar with time scales of $\sim$50 days at two frequencies. A fourth flux enhancement had a variability timescale approximately twice as long. We found that 225\,GHz enhancements led the 15\,GHz enhancements by a range of 7 to 30 days in the time delay analysis. We found the fractional variability did not change with frequency between 43 and 86\,GHz.
We could reliably measure the turnover frequency, $\nu_{\rm c}$, of the core of the source in three epochs. This was measured to be in a range from 27 to 50\,GHz and a flux density at the turnover frequency, $S_{\rm m}$, ranging from 3-6\,Jy. The derived SSA magnetic fields, $B_{\rm SSA}$, are in a range from $0.157\pm0.104$ to $0.255\pm0.146$ mG. We estimated the equipartition magnetic field strengths to be in a range from  $0.95\pm0.15$ to $1.93\pm0.30$ mG.
The equipartition magnetic field strengths are up to a factor of 10 higher than the values of $B_{\rm SSA}$. 
We conclude that the downstream jet may be more particle energy dominated.

\end{abstract}


\section{Introduction}\label{sec:intro}
Blazars are a subclass of radio-loud active galactic nuclei (AGN) which have a relativistic jet oriented toward the observer's line of sight.
Blazars typically have a small viewing angle of $<$ 20$^\circ$ \citep{Angel+80,Urry+95}.
The high Doppler factors associated with these small viewing angles shortens the observed flux density variation time scales and increases the brightness of observed flux densities.
This variability as well as polarized emission has been observed on various time scales from days to years across the whole electromagnetic spectrum \citep{Rani+13,Lee+16a,Lee+16b,Lee+17a,Lee+17b,Algaba+18a}.

 Simultaneous multi-frequency Very Long Baseline Interferometry (VLBI) observations allow us to better understand the relativistic jets of AGN that emit synchrotron radiation. It enables us to investigate magnetic field strengths which is known to play an important role in jet formation, acceleration, and collimation \citep{Begelman+84}.
 We can do this by measuring the turnover frequency and the flux density, and the size of the source at the turnover frequency \citep{Rybicki+86}. In previous studies using the Interferometric Monitoring of GAmma-ray Bright AGNs (iMOGABA) program \citep{Lee+13,Lee+16a,Lee+17a,Algaba+18a,Algaba+18b,Kim+18,Park+19} which is a part of the Korean VLBI Network (KVN) key science programs, magnetic field strengths of a few mG were found in the SSA regions of the relativistic jets in S5 0716+714~\citep{Lee+17a} and 1633+382~\citep{Algaba+18a}.

OJ~287 is a well-known BL Lacertae-type blazar with a redshift z=0.306 \citep{Sitko+85,Nilsson+10}.
Quasi-periodic flares with a period of 12 years at optical wavelengths have been detected \citep{Villforth+10}. This is thought to be caused by a binary supermassive black hole system where a secondary black hole penetrates the accretion disk of the primary system \citep{Sillanpaa+88}. If OJ~287 is modeled as a such binary system, a black hole mass of 1.8$\times10^{10}\rm M_{\odot}$ can be derived for the primary black hole \citep{Valtonen+08}. 
\cite {Valtonen+13} suggested helical jet models to explain the source variability at optical and radio bands. 
\cite{Myserlis+18} introduced a helical jet model where a polarized emission component is propagating on a helical trajectory within a bent jet, for explaining the flux density and polarization variability at radio and optical bands.
This can be used to explain the flux density variability.
\citet{Britzen+18} interpreted the flux density variation at radio bands in OJ~287 within the context of a jet precession model finding a period of $\sim$22 years. \cite{Valtonen+06} also applied the jet precession model to interpret the source variability.  

VLBI observations have been conducted to study the milli-arcsecond (mas) scale structures in the central regions of OJ~287, including the bright core with a one-sided jet in Very Long Baseline Array (VLBA) images at 43~GHz \citep{Agudo+11}. 
A quasi-stationary feature within 2 mas of the core associated with a bent jet, was analyzed using Global mm-VLBI Array (GMVA) observations at 86~GHz~\citep{Hodgson+17}.
The spectral index showed different behavior between the core and stationary component at mm-wavelengths. The spectral index of the core at 15-43\,GHz and 43-86\,GHz was inverted. When the core was flaring, the spectral index changed to become steep \citep{Hodgson+17}.

In this paper, we report results from simultaneous multiple frequency VLBI observations of OJ\,287 using the KVN at 22, 43, 86, and 129 GHz, from 2013 January until 2016 December.
The paper is organized as follows. In Section \ref{sec:obs}, we present the description of the observations for both the KVN data and all archival data used in this paper. Section \ref{sec:result} and \ref{sec:analysis} describes results and various analyses of the flux density variability at multiple frequencies, respectively and Section \ref{sec:discussion} includes discussion on the results and analysis. Finally, we summarize this paper in Section \ref{sec:summary}.
We use luminosity distance $D_{L}$=1630\,$\rm Mpc$ at a redshift of z=0.306, assuming $H_{\circ}$ = 68\,$\rm km\,\rm s^{-1}\,\rm Mpc^{-1}$, $\Omega_{\lambda} = 0.698$ and $\Omega_{m}$ = 0.302 \citep{Spergel+15,Hodgson+16}. Spectral index $\alpha$ is defined as $S_{\nu} \propto \nu^{\alpha}$.

\section{Observations and data reduction}\label{sec:obs}
\subsection{KVN observations}\label{subsec:kvn_data}

OJ~287 was observed by the KVN, which is a 500\,km-scale VLBI network consisting of three 21-m radio telescopes located in Seoul (KVN Yonsei), Ulsan (KVN Ulsan), and Jeju (KVN Tamna), Korea.
The observations were conducted at 22, 43, 86, and 129\,GHz bands simultaneously with a mean cadence of roughly a month since December 2012 except for maintenance seasons from June to August.

\startlongtable
\begin{deluxetable*}{llcccrrrrrr}
\tabletypesize{\small}
\tablecaption{Image parameters\label{table:image_para}}
\tablewidth{0pt}
\tablehead{
\colhead{Epoch} & 
\colhead{MJD} & 
\colhead{Band} & 
\colhead{$B_{\rm maj}$} &
\colhead{$B_{\rm min}$} & 
\colhead{$B_{\rm PA}$} & 
\colhead{$S_{\rm KVN}$} &
\colhead{$S_{\rm p}$} &
\colhead{$\sigma$} &
\colhead{$D$} &
\colhead{$\xi_{\rm r}$}\\ 
\colhead{(1)} & 
\colhead{(2)} & 
\colhead{(3)} &
\colhead{(4)} & 
\colhead{(5)} & 
\colhead{(6)} &
\colhead{(7)} &
\colhead{(8)} &
\colhead{(9)} &
\colhead{(10)} &
\colhead{(11)}
}
\startdata
2013 Jan 16&     56308 &K   &      5.396 &      3.234 &  $-$81.8 &       3.495 &       3.416 &          22 &    155 &  0.64 \\
           &           &Q   &      2.715 &      1.603 &  $-$80.0 &       3.792 &       3.616 &          26 &    141 &  0.71 \\
           &           &W   &      1.365 &      0.765 &  $-$69.7 &       2.692 &       2.712 &          45 &     60 &  0.60 \\
           &           &D   &      1.388 &      0.764 &  $-$69.6 &       2.520 &       2.599 &          75 &     35 &  0.55 \\
2013 Mar 28&     56379 &K   &      9.785 &      3.131 &  $-$45.6 &       5.566 &       5.579 &          24 &    237 &  0.46 \\
           &           &D   &      1.610 &      0.498 &  $-$50.4 &       2.320 &       2.529 &         363 &      7 &  0.42 \\
2013 Apr 11&     56393 &K   &      5.616 &      3.161 &  $-$77.7 &       5.095 &       4.963 &          17 &    291 &  0.61 \\
           &           &Q   &      2.765 &      1.523 &  $-$79.6 &       4.525 &       4.238 &          17 &    255 &  0.55 \\
           &           &W   &      1.556 &      0.708 &  $-$66.8 &       3.614 &       2.555 &          59 &     44 &  0.60 \\
           &           &D   &      1.013 &      0.502 &  $-$64.8 &       2.874 &       1.594 &          64 &     25 &  0.60 \\
2013 May 07&     56419 &K   &      5.344 &      3.245 &  $-$78.3 &       4.029 &       3.980 &          10 &    407 &  0.60 \\
           &           &Q   &      2.641 &      1.728 &  $-$80.6 &       3.385 &       3.237 &          15 &    210 &  0.67 \\
           &           &W   &      1.377 &      0.753 &  $-$75.3 &       2.686 &       2.401 &          31 &     77 &  0.59 \\
           &           &D   &      0.920 &      0.513 &  $-$71.0 &       2.047 &       1.852 &          37 &     50 &  0.68 \\
2013 Sep 24&     56559 &K   &      7.611 &      2.996 &  $-$55.4 &       4.189 &       4.048 &          24 &    168 &  0.60 \\
           &           &Q   &      2.912 &      1.547 &  $-$64.7 &       2.867 &       2.865 &          33 &     88 &  0.67 \\
2013 Oct 15&     56580 &K   &      6.199 &      3.720 &       71.2 &       4.071 &       3.841 &          13 &    296 &  0.62 \\
           &           &Q   &      2.893 &      1.836 &       80.1 &       3.323 &       2.968 &          42 &     71 &  0.66 \\
           &           &W   &      1.447 &      0.781 &       82.0 &       2.602 &       2.442 &          38 &     64 &  0.63 \\
           &           &D   &      1.459 &      0.801 &       84.4 &       2.633 &       2.459 &          36 &     68 &  0.60 \\
2013 Nov 19&     56615 &K   &      6.286 &      3.074 &  $-$65.7 &       4.703 &       4.688 &          43 &    109 &  0.61 \\
           &           &Q   &      3.233 &      1.497 &  $-$67.7 &       3.873 &       3.693 &          17 &    217 &  0.60 \\
           &           &W   &      1.784 &      0.698 &  $-$62.9 &       2.863 &       2.615 &          18 &    145 &  0.66 \\
           &           &D   &      1.062 &      0.520 &  $-$55.5 &       1.925 &       1.803 &          22 &     82 &  0.57 \\
2013 Dec 24&     56650 &K   &      5.366 &      3.079 &  $-$79.7 &       3.181 &       3.086 &          11 &    282 &  0.77 \\
           &           &Q   &      2.612 &      1.600 &  $-$79.7 &       2.501 &       2.414 &          13 &    191 &  0.87 \\
           &           &W   &      1.397 &      0.712 &  $-$80.1 &       1.898 &       1.812 &          15 &    124 &  0.77 \\
           &           &D   &      1.021 &      0.442 &  $-$75.6 &       1.362 &       1.340 &          25 &     54 &  0.63 \\
2014 Jan 27&     56684 &K   &      5.782 &      3.049 &  $-$68.8 &       2.765 &       2.761 &          16 &    172 &  0.72 \\
           &           &Q   &      2.747 &      1.602 &  $-$68.0 &       2.325 &       2.289 &          18 &    128 &  0.71 \\
2014 Feb 28&     56716 &K   &      6.034 &      2.871 &  $-$66.6 &       3.110 &       2.971 &          13 &    232 &  0.47 \\
           &           &Q   &      3.039 &      1.434 &  $-$65.9 &       2.861 &       2.644 &           8 &    334 &  0.67 \\
           &           &W   &      2.097 &      0.605 &  $-$72.1 &       2.223 &       2.119 &          14 &    155 &  0.60 \\
           &           &D   &      1.481 &      0.402 &  $-$71.5 &       1.766 &       1.619 &          31 &     52 &  0.53 \\
2014 Mar 22&     56738 &K   &      5.515 &      3.089 &  $-$68.4 &       2.851 &       2.776 &           9 &    324 &  0.76 \\
           &           &Q   &      2.745 &      1.562 &  $-$66.4 &       2.773 &       2.687 &          15 &    175 &  0.72 \\
           &           &W   &      1.495 &      0.718 &  $-$69.7 &       2.551 &       2.478 &          17 &    148 &  0.66 \\
           &           &D   &      1.016 &      0.484 &  $-$67.2 &       1.807 &       1.659 &          67 &     25 &  0.66 \\
2014 Apr 22&     56769 &K   &      5.388 &      3.131 &  $-$76.2 &       4.278 &       4.171 &          19 &    224 &  0.57 \\
           &           &Q   &      2.838 &      1.544 &  $-$71.3 &       4.877 &       4.778 &          34 &    141 &  0.62 \\
           &           &W   &      1.461 &      0.725 &  $-$71.9 &       4.997 &       4.874 &          48 &    103 &  0.64 \\
           &           &D   &      0.982 &      0.493 &  $-$68.4 &       3.793 &       3.700 &          85 &     44 &  0.68 \\
2014 Sep 01&     56902 &K   &      5.399 &      3.250 &  $-$80.4 &       4.843 &       4.743 &          18 &    267 &  0.68 \\
           &           &Q   &      2.892 &      1.531 &  $-$74.6 &       4.465 &       4.261 &          22 &    196 &  0.69 \\
           &           &W   &      1.408 &      0.731 &  $-$79.1 &       3.477 &       3.357 &          37 &     90 &  0.66 \\
           &           &D   &      1.408 &      0.735 &  $-$76.7 &       3.536 &       3.394 &          35 &     98 &  0.63 \\
2014 Oct 29&     56960 &K   &      5.307 &      3.220 &  $-$79.2 &       5.731 &       5.535 &          18 &    303 &  0.89 \\
           &           &Q   &      2.636 &      1.691 &  $-$85.0 &       5.023 &       4.679 &          22 &    211 &  0.60 \\
           &           &W   &      1.464 &      0.659 &  $-$79.6 &       3.317 &       3.138 &          41 &     77 &  0.58 \\
           &           &D   &      1.564 &      0.493 &       58.4 &       1.387 &       1.352 &          72 &     19 &  0.44 \\
2014 Nov 28&     56990 &K   &      5.717 &      2.847 &  $-$81.7 &       5.028 &       5.021 &          51 &     99 &  0.76 \\
           &           &Q   &      2.809 &      1.467 &  $-$82.5 &       4.485 &       4.433 &          43 &    104 &  0.58 \\
           &           &W   &      1.527 &      0.676 &  $-$73.7 &       3.493 &       3.351 &          43 &     78 &  0.65 \\
2014 Dec 25&     57017 &K   &      5.503 &      3.005 &  $-$77.4 &       5.668 &       5.620 &          23 &    249 &  0.65 \\
2015 Feb 23&     57076 &K   &      5.331 &      3.149 &  $-$81.2 &       7.144 &       7.089 &          15 &    464 &  0.57 \\
           &           &Q   &      2.661 &      1.584 &  $-$78.0 &       5.825 &       5.746 &          13 &    456 &  0.61 \\
           &           &W   &      1.419 &      0.732 &  $-$77.6 &       4.389 &       4.278 &          19 &    229 &  0.73 \\
           &           &D   &      0.934 &      0.492 &  $-$83.5 &       3.291 &       1.414 &          22 &     64 &  0.64 \\
2015 Mar 26&     57107 &K   &      5.534 &      3.030 &  $-$74.1 &       5.162 &       5.065 &          11 &    470 &  0.70 \\
           &           &Q   &      2.804 &      1.517 &  $-$73.3 &       4.703 &       4.624 &          11 &    429 &  0.70 \\
           &           &W   &      1.465 &      0.721 &  $-$73.0 &       4.189 &       3.963 &          34 &    118 &  0.64 \\
           &           &D   &      0.956 &      0.507 &  $-$72.7 &       3.348 &       3.059 &          43 &     71 &  0.61 \\
2015 Apr 30&     57142 &K   &      5.412 &      3.119 &       83.8 &       4.958 &       4.895 &          42 &    118 &  0.80 \\
           &           &W   &      1.518 &      0.708 &       82.7 &       3.483 &  3.261 &          35 &     93 &  0.64 \\
           &           &D   &      1.001 &      0.465 &  $-$88.9 &       2.464 &       2.511 &          49 &     51 &  0.68 \\
2015 Sep 24&     57289 &K   &      5.280 &      3.103 &  $-$84.8 &       3.377 &       3.344 &          25 &    133 &  0.88 \\
           &           &Q   &      2.659 &      1.480 &  $-$84.7 &       2.472 &       2.452 &          24 &    104 &  0.67 \\
           &           &W   &      1.458 &      0.701 &       86.8 &       1.831 &       1.753 &          27 &     65 &  0.60 \\
2015 Oct 23&     57318 &K   &      5.660 &      3.079 &  $-$89.2 &       2.578 &       2.468 &          15 &    166 &  0.66 \\
           &           &Q   &      2.879 &      1.489 &  $-$87.2 &       2.267 &       2.076 &          16 &    133 &  0.69 \\
           &           &W   &      1.691 &      0.702 &       85.5 &       2.440 &       2.343 &          39 &     60 &  0.57 \\
2015 Nov 30&     57356 &K   &      5.398 &      3.101 &  $-$78.9 &       3.240 &       3.191 &          19 &    166 &  0.87 \\
           &           &Q   &      2.681 &      1.510 &  $-$82.2 &       3.151 &       3.150 &          14 &    222 &  0.59 \\
           &           &W   &      1.358 &      0.745 &  $-$85.4 &       3.084 &       2.932 &          32 &     90 &  0.53 \\
           &           &D   &      0.911 &      0.502 &  $-$81.0 &       2.172 &       2.150 &          36 &     60 &  0.73 \\
2015 Dec 28&     57384 &K   &      5.257 &      3.138 &  $-$86.6 &       3.702 &       3.699 &          41 &     91 &  0.73 \\
           &           &Q   &      2.761 &      1.513 &  $-$82.0 &       3.635 &       3.631 &          24 &    153 &  0.58 \\
           &           &W   &      1.460 &      0.718 &       86.9 &       3.093 &       3.008 &          39 &     76 &  0.63 \\
           &           &D   &      0.934 &      0.497 &  $-$86.9 &       1.809 &       1.599 &          34 &     47 &  0.58 \\
2016 Jan 13&     57400 &K   &      5.171 &      3.284 &       80.4 &       3.227 &       3.224 &          28 &    117 &  0.58 \\
           &           &Q   &      2.694 &      1.529 &       89.2 &       2.997 &       2.983 &          90 &     33 &  0.61 \\
           &           &W   &      1.386 &      0.767 &       80.9 &       2.519 &       2.283 &          25 &     91 &  0.51 \\
           &           &D   &      0.868 &      0.547 &       80.5 &       1.696 &       1.630 &          33 &     49 &  0.58 \\
2016 Feb 11&     57429 &K   &      5.685 &      3.678 &       60.2 &       3.131 &       3.122 &          48 &     65 &  0.61 \\
           &           &Q   &      2.835 &      1.838 &       58.9 &       3.433 &       3.062 &          53 &     57 &  0.62 \\
2016 Mar 01&     57448 &K   &      5.343 &      3.095 &  $-$83.1 &       4.020 &       3.964 &          21 &    192 &  0.80 \\
           &           &Q   &      2.686 &      1.544 &  $-$84.6 &       4.202 &       4.111 &          20 &    207 &  0.74 \\
           &           &W   &      1.373 &      0.750 &  $-$87.9 &       3.328 &       3.261 &          34 &     95 &  0.68 \\
           &           &D   &      0.896 &      0.540 &  $-$75.3 &       2.407 &       2.363 &          38 &     63 &  0.71 \\
2016 Apr 25&     57503 &K   &      5.312 &      3.308 &  $-$78.1 &       5.461 &       5.373 &          61 &     88 &  0.72 \\
           &           &Q   &      2.673 &      1.627 &  $-$78.1 &       5.243 &       4.908 &          28 &    176 &  0.65 \\
           &           &W   &      1.350 &      0.780 &  $-$77.0 &       3.925 &       3.832 &          38 &    101 &  0.66 \\
2016 Aug 23&     57623 &K   &      5.918 &      3.101 &  $-$66.7 &       8.440 &       8.399 &          50 &    169 &  0.69 \\
           &           &W   &      1.541 &      0.762 &  $-$59.5 &       4.889 &       4.916 &          90 &     55 &  0.68 \\
           &           &D   &      1.019 &      0.543 &  $-$54.5 &       3.753 &       3.751 &         166 &     23 &  0.69 \\
2016 Oct 18&     57680 &K   &      5.860 &      2.907 &       82.7 &       7.122 &       6.948 &          39 &    180 &  0.73 \\
           &           &Q   &      3.180 &      1.355 &       82.6 &       5.657 &       5.480 &          29 &    190 &  0.83 \\
           &           &W   &      1.836 &      0.627 &       81.4 &       4.418 &       4.412 &          12 &    383 &  0.60 \\
           &           &D   &      1.205 &      0.426 &       83.5 &       3.147 &       3.149 &          42 &     75 &  0.67 \\
2016 Nov 27&     57720 &K   &      5.499 &      3.472 &       72.3 &       6.316 &       6.370 &          41 &    157 &  0.81 \\
           &           &Q   &      3.376 &      1.502 &  $-$63.2 &       5.220 &       4.653 &          35 &    131 &  0.57 \\
           &           &W   &      1.703 &      0.738 &  $-$65.2 &       3.917 &       3.784 &          58 &     65 &  0.54 \\
2016 Dec 28&     57751 &K   &      5.136 &      3.257 &  $-$86.2 &       8.104 &       7.729 &          29 &    269 &  0.57 \\
           &           &Q   &      2.689 &      1.561 &  $-$82.7 &       6.786 &       6.375 &          26 &    244 &  0.74 \\
           &           &W   &      1.311 &      0.794 &       84.2 &       5.909 &       4.983 &          43 &    117 &  0.67 \\
           &           &D   &      0.866 &      0.579 &  $-$78.9 &       3.684 &       3.689 &          77 &     48 &  0.67 \\
\enddata
\tablecomments{ 
Column designation: 1~-~Date; 2~-~modified Julian date; 3~-~observing frequency band: 
K~-~22~GHz band; Q~-~43~GHz band; W~-~86~GHz band; D~-~129~GHz band; 
4-6~-~restoring beam:
4~-~major axis [mas];
5~-~minor axis [mas];
6~-~position angle of the major axis [degree];
7~-~total cleaned KVN flux density [Jy];
8~-~peak flux density [Jy beam$^{-1}$];
9~-~off-source RMS in the image [mJy beam$^{-1}$];
10~-~dynamic range of the image;
11~-~quality of the residual noise in the image (i.e., ratio of the image root-mean-square noise to its mathematical expectation).
}
\end{deluxetable*}
\clearpage
The angular resolution of the KVN is 6 mas at 22\,GHz, 3 mas at 43\,GHz, 1.5 mas at 86\,GHz, and 1  mas at 129\,GHz.
We used data obtained in 31 epochs in the period from 2013 January 16 (MJD 56308) until 2016 December 28 (MJD 57750).
In each epoch, the source was observed with up to eight  5 minute scans per 24 hours. The observations were recorded in left circular polarization mode at a recording rate of 1\,Gbps. 
The observing frequencies are 21.700-21.764\,GHz, 43.400-43.464\,GHz, 86.800-86.864\,GHz, and 129.300-129.364\,GHz, with a total bandwidth of 256\,MHz (thus 64\,MHz in each band). 
Detailed descriptions of the iMOGABA observations are reported in \citet{Lee+16a} and \citet{Lee+17a}.

\startlongtable
\begin{deluxetable*}{llcccrr}
\tabletypesize{\small}
\tablecaption{Modelfit parameters for the center of images\label{table:modelfit}}
\tablewidth{0pt}
\tablehead{
\colhead{Epoch} & 
\colhead{MJD} & 
\colhead{Band} & 
\colhead{$S_{\rm tot}$} &
\colhead{$S_{\rm peak}$} & 
\colhead{$d$} & 
\colhead{$T_{\rm b}$} \\ 
\colhead{(1)} & 
\colhead{(2)} & 
\colhead{(3)} &
\colhead{(4)} & 
\colhead{(5)} & 
\colhead{(6)} &
\colhead{(7)} }
\startdata
2013 Jan 16  &       56308   &  K   &    3.54$\pm$ 0.05 &    3.40$\pm$ 0.04 &   0.85$\pm$ 0.01    &     0.90$\pm$  0.02 \\
             &               &  Q   &    3.73$\pm$ 0.09 &    3.60$\pm$ 0.06 &   0.40$\pm$ 0.01    &     4.41$\pm$  0.15 \\
             &               &  W   &    2.68$\pm$ 0.55 &    2.71$\pm$ 0.39 &  $<${\it  0.14}     &     $>${\it  25.86} \\
             &               &  D   &    2.56$\pm$ 0.57 &    2.60$\pm$ 0.41 &  $<${\it  0.15}     &     $>${\it  20.14} \\
2013 Mar 28  &       56379   &  K   &    5.58$\pm$ 0.03 &    5.58$\pm$ 0.02 &   0.14$\pm$ 0.00    &     55.09$\pm$  0.39 \\
             &               &  D   &    6.25$\pm$ 0.71 &    2.83$\pm$ 0.29 &   1.29$\pm$ 0.13    &     0.70$\pm$  0.14 \\
2013 Apr 11  &       56393   &  K   &    5.05$\pm$ 0.10 &    4.95$\pm$ 0.07 &   0.60$\pm$ 0.01    &     2.63$\pm$  0.08 \\
             &               &  Q   &    4.42$\pm$ 0.17 &    4.21$\pm$ 0.12 &   0.45$\pm$ 0.01    &     4.13$\pm$  0.24 \\
             &               &  W   &    4.03$\pm$ 0.78 &    2.36$\pm$ 0.39 &   0.74$\pm$ 0.12    &     1.38$\pm$  0.46 \\
             &               &  D   &    4.18$\pm$ 2.25 &    1.30$\pm$ 0.66 &   0.80$\pm$ 0.41    &     1.21$\pm$  1.24 \\
2013 May 07  &       56419   &  K   &    3.96$\pm$ 0.46 &    3.98$\pm$ 0.33 &  $<${\it  0.32}     &     $>${\it   7.15} \\
             &               &  Q   &    3.34$\pm$ 0.11 &    3.22$\pm$ 0.08 &   0.42$\pm$ 0.01    &     3.53$\pm$  0.17 \\
             &               &  W   &    2.90$\pm$ 0.13 &    2.38$\pm$ 0.09 &   0.45$\pm$ 0.02    &     2.63$\pm$  0.19 \\
             &               &  D   &    2.23$\pm$ 0.09 &    1.83$\pm$ 0.05 &   0.30$\pm$ 0.01    &     4.49$\pm$  0.26 \\
2013 Sep 24  &       56559   &  K   &    4.14$\pm$ 0.08 &    4.01$\pm$ 0.06 &   0.75$\pm$ 0.01    &     1.36$\pm$  0.04 \\
             &               &  Q   &    2.87$\pm$ 0.18 &    2.86$\pm$ 0.13 &   0.16$\pm$ 0.01    &     21.87$\pm$  1.95 \\
2013 Oct 15  &       56580   &  K   &    4.03$\pm$ 0.12 &    3.87$\pm$ 0.08 &   0.98$\pm$ 0.02    &     0.77$\pm$  0.03 \\
             &               &  Q   &    3.57$\pm$ 0.25 &    2.97$\pm$ 0.16 &   1.03$\pm$ 0.05    &     0.63$\pm$  0.07 \\
             &               &  W   &    2.83$\pm$ 0.15 &    2.43$\pm$ 0.10 &   0.41$\pm$ 0.02    &      3.11$\pm$  0.25 \\
             &               &  D   &    2.88$\pm$ 0.14 &    2.45$\pm$ 0.09 &   0.43$\pm$ 0.02    &       2.87$\pm$  0.21 \\
2013 Nov 19  &       56615   &  K   &    4.75$\pm$ 0.06 &    4.69$\pm$ 0.04 &   0.49$\pm$ 0.00    &         3.72$\pm$  0.07 \\
             &               &  Q   &    3.84$\pm$ 0.06 &    3.68$\pm$ 0.04 &   0.43$\pm$ 0.01    &        3.90$\pm$  0.09 \\
             &               &  W   &    2.91$\pm$ 0.11 &    2.58$\pm$ 0.07 &   0.35$\pm$ 0.01    &        4.48$\pm$  0.26 \\
             &               &  D   &    1.94$\pm$ 0.11 &    1.81$\pm$ 0.07 &   0.19$\pm$ 0.01    &       10.06$\pm$  0.82 \\
2013 Dec 24  &       56650   &  K   &    3.06$\pm$ 0.42 &    3.08$\pm$ 0.30 &  $<${\it  0.37}     &        $>${\it   4.21} \\
             &               &  Q   &    2.43$\pm$ 0.04 &    2.41$\pm$ 0.03 &   0.20$\pm$ 0.00    &       11.36$\pm$  0.25 \\
             &               &  W   &    1.96$\pm$ 0.08 &    1.81$\pm$ 0.05 &   0.27$\pm$ 0.01    &        4.90$\pm$  0.28 \\
             &               &  D   &    1.32$\pm$ 0.30 &    1.34$\pm$ 0.22 &  $<${\it  0.10}     &      $>${\it  22.99} \\
2014 Jan 27  &       56684   &  K   &    2.75$\pm$ 0.28 &    2.76$\pm$ 0.20 &  $<${\it  0.28}     &      $>${\it   6.55} \\
             &               &  Q   &    2.32$\pm$ 0.02 &    2.28$\pm$ 0.01 &   0.27$\pm$ 0.00    &        5.81$\pm$  0.06 \\
2014 Feb 28  &       56716   &  K   &    2.93$\pm$ 0.48 &    2.96$\pm$ 0.34 &  $<${\it  0.46}     &        $>${\it   2.62} \\
             &               &  Q   &    2.59$\pm$ 0.47 &    2.61$\pm$ 0.33 &  $<${\it  0.25}     &        $>${\it   7.55} \\
             &               &  W   &    2.09$\pm$ 0.38 &    2.11$\pm$ 0.27 &  $<${\it  0.14}     &        $>${\it  20.90} \\
             &               &  D   &    1.87$\pm$ 0.09 &    1.59$\pm$ 0.06 &   0.24$\pm$ 0.01    &          6.02$\pm$  0.45 \\
2014 Mar 22  &       56738   &  K   &    2.77$\pm$ 0.02 &    2.77$\pm$ 0.02 &   0.10$\pm$ 0.00    &         49.30$\pm$  0.62 \\
             &               &  Q   &    2.67$\pm$ 0.34 &    2.68$\pm$ 0.24 &  $<${\it  0.17}     &      $>${\it  16.25} \\
             &               &  W   &    2.45$\pm$ 0.52 &    2.47$\pm$ 0.37 &  $<${\it  0.15}     &        $>${\it  21.05} \\
             &               &  D   &    1.69$\pm$ 0.12 &    1.64$\pm$ 0.08 &   0.11$\pm$ 0.01    &         24.02$\pm$  2.40 \\
2014 Apr 22  &       56769   &  K   &    4.14$\pm$ 0.45 &    4.16$\pm$ 0.32 &  $<${\it  0.30}     &          $>${\it   8.73} \\
             &               &  Q   &    4.65$\pm$ 1.49 &    4.76$\pm$ 1.07 &  $<${\it  0.45}     &       $>${\it   4.30} \\
             &               &  W   &    5.22$\pm$ 0.09 &    4.86$\pm$ 0.06 &   0.26$\pm$ 0.00    &      13.86$\pm$  0.36 \\
             &               &  D   &    3.98$\pm$ 0.10 &    3.69$\pm$ 0.07 &   0.18$\pm$ 0.00    &      21.78$\pm$  0.84 \\
2014 Sep 01  &       56902   &  K   &    4.96$\pm$ 0.04 &    4.74$\pm$ 0.02 &   0.91$\pm$ 0.00    &    1.11$\pm$  0.01 \\
             &               &  Q   &    4.50$\pm$ 0.10 &    4.24$\pm$ 0.07 &   0.50$\pm$ 0.01    &    3.29$\pm$  0.11 \\
             &               &  W   &    3.78$\pm$ 0.16 &    3.35$\pm$ 0.10 &   0.34$\pm$ 0.01    &    5.98$\pm$  0.37 \\
             &               &  D   &    3.79$\pm$ 0.13 &    3.38$\pm$ 0.09 &   0.33$\pm$ 0.01    &    6.33$\pm$  0.32 \\
2014 Oct 29  &       56960   &  K   &    5.62$\pm$ 0.08 &    5.49$\pm$ 0.05 &   0.65$\pm$ 0.01    &   2.46$\pm$  0.05 \\
             &               &  Q   &    4.86$\pm$ 0.07 &    4.66$\pm$ 0.05 &   0.45$\pm$ 0.00    &   4.55$\pm$  0.10 \\
             &               &  W   &    3.46$\pm$ 0.22 &    3.14$\pm$ 0.15 &   0.29$\pm$ 0.01    &   7.83$\pm$  0.73 \\
             &               &  D   &    1.33$\pm$ 0.32 &    1.35$\pm$ 0.23 &  $<${\it  0.14}     &  $>${\it  12.61} \\
2014 Nov 28  &       56990   &  K   &    5.62$\pm$ 0.99 &    5.02$\pm$ 0.66 &   1.20$\pm$ 0.16    &        0.72$\pm$  0.19 \\
             &               &  Q   &    4.43$\pm$ 0.18 &    4.43$\pm$ 0.13 &  $<${\it  0.06}     & $>${\it 267.51} \\
             &               &  W   &    3.96$\pm$ 0.05 &    3.34$\pm$ 0.04 &   0.39$\pm$ 0.00    &        4.87$\pm$  0.10 \\
2014 Dec 25  &       57017   &  K   &    5.62$\pm$ 0.04 &    5.61$\pm$ 0.02 &   0.16$\pm$ 0.00    &   39.17$\pm$  0.35 \\
2015 Feb 23  &       57076   &  K   &    7.07$\pm$ 0.71 &    7.09$\pm$ 0.50 &  $<${\it  0.27}     &  $>${\it  17.39} \\
             &               &  Q   &    5.76$\pm$ 0.04 &    5.74$\pm$ 0.03 &   0.12$\pm$ 0.00    &        79.23$\pm$  0.78 \\
             &               &  W   &    4.46$\pm$ 0.12 &    4.28$\pm$ 0.08 &   0.20$\pm$ 0.00    &   19.66$\pm$  0.75 \\
             &               &  D   &    3.84$\pm$ 1.82 &    1.07$\pm$ 0.49 &   1.19$\pm$ 0.54    &   0.50$\pm$  0.46 \\
2015 Mar 26  &       57107   &  K   &    5.13$\pm$ 0.06 &    5.05$\pm$ 0.04 &   0.50$\pm$ 0.00    &       3.85$\pm$  0.06 \\
             &               &  Q   &    4.72$\pm$ 0.05 &    4.62$\pm$ 0.04 &   0.30$\pm$ 0.00    &               9.77$\pm$  0.16 \\
             &               &  W   &    4.47$\pm$ 0.16 &    3.94$\pm$ 0.11 &   0.35$\pm$ 0.01    &               6.83$\pm$  0.38 \\
             &               &  D   &    3.30$\pm$ 0.09 &    3.01$\pm$ 0.06 &   0.21$\pm$ 0.00    &               14.26$\pm$  0.58 \\
2015 Apr 30  &       57142   &  K   &    4.83$\pm$ 0.73 &    4.88$\pm$ 0.52 &  $<${\it  0.41}     &              $>${\it   5.23} \\
             &               &  W   &    3.38$\pm$ 0.14 &    3.22$\pm$ 0.09 &   0.22$\pm$ 0.01    &              13.31$\pm$  0.78 \\
             &               &  D   &    2.55$\pm$ 0.04 &    2.51$\pm$ 0.03 &   0.08$\pm$ 0.00    &      68.62$\pm$  1.61 \\
2015 Sep 24  &       57289   &  K   &    3.30$\pm$ 0.52 &    3.34$\pm$ 0.37 &  $<${\it  0.42}     &             $>${\it   3.43} \\
             &               &  Q   &    2.47$\pm$ 0.02 &    2.45$\pm$ 0.02 &   0.17$\pm$ 0.00    &             16.20$\pm$  0.21 \\
             &               &  W   &    2.01$\pm$ 0.09 &    1.75$\pm$ 0.06 &   0.36$\pm$ 0.01    &                2.90$\pm$  0.19 \\
2015 Oct 23  &       57318   &  K   &    2.43$\pm$ 0.31 &    2.45$\pm$ 0.22 &  $<${\it  0.35}     &             $>${\it   3.67} \\
             &               &  Q   &    2.21$\pm$ 0.07 &    2.04$\pm$ 0.05 &   0.56$\pm$ 0.01    &              1.29$\pm$  0.06 \\
             &               &  W   &    2.77$\pm$ 0.07 &    2.34$\pm$ 0.05 &   0.41$\pm$ 0.01    &            3.09$\pm$  0.12 \\
2015 Nov 30  &       57356   &  K   &    3.14$\pm$ 0.50 &    3.18$\pm$ 0.35 &  $<${\it  0.43}     &              $>${\it   3.11} \\
             &               &  Q   &    3.14$\pm$ 0.31 &    3.15$\pm$ 0.22 &  $<${\it  0.13}     &              $>${\it  32.90} \\
             &               &  W   &    3.11$\pm$ 0.14 &    2.92$\pm$ 0.10 &   0.25$\pm$ 0.01    &            9.51$\pm$  0.63 \\
             &               &  D   &    2.06$\pm$ 0.90 &    2.15$\pm$ 0.65 &  $<${\it  0.20}     &             $>${\it   9.79} \\
2015 Dec 28  &       57384   &  K   &    3.69$\pm$ 0.35 &    3.70$\pm$ 0.25 &  $<${\it  0.26}     &             $>${\it  10.13} \\
             &               &  Q   &    3.75$\pm$ 0.30 &    3.63$\pm$ 0.21 &   0.34$\pm$ 0.02    &               5.90$\pm$  0.67 \\
             &               &  W   &    3.50$\pm$ 0.13 &    3.00$\pm$ 0.09 &   0.38$\pm$ 0.01    &             4.42$\pm$  0.25 \\
             &               &  D   &    1.90$\pm$ 0.19 &    1.58$\pm$ 0.12 &   0.28$\pm$ 0.02    &                4.33$\pm$  0.66 \\
2016 Jan 13  &       57400   &  K   &    3.20$\pm$ 0.49 &    3.22$\pm$ 0.35 &  $<${\it  0.42}     &              $>${\it   3.41} \\
             &               &  Q   &    3.56$\pm$ 0.28 &    2.98$\pm$ 0.18 &   0.85$\pm$ 0.05    &             0.91$\pm$  0.11 \\
             &               &  W   &    2.94$\pm$ 0.27 &    2.28$\pm$ 0.16 &   0.52$\pm$ 0.04    &              2.05$\pm$  0.29 \\
             &               &  D   &    1.70$\pm$ 0.09 &    1.63$\pm$ 0.06 &   0.14$\pm$ 0.01    &         14.98$\pm$  1.15 \\
2016 Feb 11  &       57429   &  K   &    3.30$\pm$ 0.09 &    3.12$\pm$ 0.06 &   1.11$\pm$ 0.02    &            0.49$\pm$  0.02 \\
             &               &  Q   &    4.22$\pm$ 0.20 &    3.20$\pm$ 0.12 &   1.25$\pm$ 0.05    &               0.50$\pm$  0.04 \\
2016 Mar 01  &       57448   &  K   &    3.91$\pm$ 0.57 &    3.95$\pm$ 0.40 &  $<${\it  0.39}     &             $>${\it   4.67} \\
             &               &  Q   &    4.39$\pm$ 0.04 &    4.11$\pm$ 0.02 &   0.53$\pm$ 0.00    &              2.91$\pm$  0.04 \\
             &               &  W   &    3.50$\pm$ 0.08 &    3.26$\pm$ 0.06 &   0.27$\pm$ 0.00    &             9.02$\pm$  0.32 \\
             &               &  D   &    2.30$\pm$ 0.68 &    2.35$\pm$ 0.48 &  $<${\it  0.14}     &              $>${\it  23.04} \\
2016 Apr 25  &       57503   &  K   &    5.28$\pm$ 1.30 &    5.37$\pm$ 0.92 &  $<${\it  0.68}     &              $>${\it   2.09} \\
             &               &  Q   &    5.40$\pm$ 0.24 &    4.89$\pm$ 0.16 &   0.67$\pm$ 0.02    &               2.25$\pm$  0.15 \\
             &               &  W   &    4.03$\pm$ 0.05 &    3.83$\pm$ 0.03 &   0.23$\pm$ 0.00    &              13.53$\pm$  0.22 \\
2016 Aug 23  &       57623   &  K   &    8.39$\pm$ 0.24 &    8.39$\pm$ 0.17 &  $<${\it  0.08}     &           $>${\it 237.56} \\
             &               &  W   &    4.94$\pm$ 0.05 &    4.92$\pm$ 0.03 &   0.06$\pm$ 0.00    &          237.71$\pm$  3.33 \\
             &               &  D   &    3.56$\pm$ 1.82 &    3.75$\pm$ 1.32 &  $<${\it  0.25}     &           $>${\it  10.40} \\
2016 Oct 18  &       57680   &  K   &    6.86$\pm$ 1.26 &    6.93$\pm$ 0.90 &  $<${\it  0.50}     &           $>${\it   4.98} \\
             &               &  Q   &    5.98$\pm$ 0.21 &    5.46$\pm$ 0.14 &   0.57$\pm$ 0.02    &            3.36$\pm$  0.18 \\
             &               &  W   &    4.53$\pm$ 0.04 &    4.41$\pm$ 0.03 &   0.14$\pm$ 0.00    &              39.85$\pm$  0.56 \\
             &               &  D   &    3.13$\pm$ 0.52 &    3.15$\pm$ 0.37 &  $<${\it  0.08}     &              $>${\it  93.26} \\
2016 Nov 27  &       57720   &  K   &    6.27$\pm$ 1.49 &    6.37$\pm$ 1.06 &  $<${\it  0.69}     &              $>${\it   2.44} \\
             &               &  Q   &    5.56$\pm$ 0.36 &    4.55$\pm$ 0.23 &   0.94$\pm$ 0.05    &                1.16$\pm$  0.12 \\
             &               &  W   &    3.88$\pm$ 0.09 &    3.78$\pm$ 0.07 &   0.17$\pm$ 0.00    &              25.44$\pm$  0.88 \\
2016 Dec 28  &       57751   &  K   &    7.54$\pm$ 1.55 &    7.62$\pm$ 1.10 &  $<${\it  0.56}     &             $>${\it   4.45} \\
             &               &  Q   &    6.85$\pm$ 0.24 &    6.29$\pm$ 0.16 &   0.60$\pm$ 0.02    &               3.56$\pm$  0.18 \\
             &               &  W   &    5.04$\pm$ 0.31 &    4.84$\pm$ 0.21 &   0.21$\pm$ 0.01    &              21.37$\pm$  1.89 \\
             &               &  D   &    3.66$\pm$ 0.44 &    3.68$\pm$ 0.31 &  $<${\it  0.06}     &            $>${\it 207.79} \\
\enddata
\tablecomments{ All the modelfit parameters are for the core component at the center of the image.
Column designation: 1~-~date; 2~-~modified Julian date;
3~-~observing frequency band: 
K~-~22~GHz band; Q~-~43~GHz band; W~-~86~GHz band; D~-~129~GHz band; 
4~-~model flux density of the component [Jy];
5~-~peak brightness of an individual component measured in the image [Jy beam$^{-1}$];
6~-~size [mas], italic numbers indicate upper limits;
7~-~measured brightness temperature [$10^9$~K], italic numbers indicate lower limits.
}
\end{deluxetable*}

\subsection{Data calibration}

The obtained data were processed with the DiFX software correlator in Daejeon, Korea \citep{Deller+07,Deller+11} and the data was calibrated following the standard procedures (phase and amplitude calibration, fringe fitting, bandpass correction) in the AIPS package from the National Radio Astronomy Observatory (NRAO). For the calibration, the KVN pipeline was used \citep{Hodgson+16}.

The visibility phases in the 22~GHz band were transferred to those in the higher frequency bands (i.e., 43, 86, and 129~GHz bands) using the pipeline in order to increase the coherence time at the higher frequency bands and hence to improve the detection sensitivity \citep{Algaba+15}.
An amplitude re-quantization loss from the digital filter was corrected by applying a factor of 1.1 to the visibility amplitude of each station~\citep{Lee+15}.
The uncertainty of the amplitude calibration is expected to be 5\,\% at 22 and 43\,GHz, 10\,\% at 86, and 30~\% at 129\,GHz.

\subsection{Imaging and Model-fitting}
After data calibration in AIPS, the KVN data were imported into the Difmap software package for imaging in the same manner as described in \citet{Lee+17a}.

The visibility data were averaged in time at 30 sec intervals at 22, 43, and 86\,GHz and 10 sec intervals at 129\,GHz, which are the typical coherence time scales of KVN observations.

We obtained a total of 106 VLBI images of OJ\,287 at four frequencies during 31 epochs over four years.
These include 31 images at 22\,GHz, 27 images at 43\,GHz, 26 images at 86\,GHz, and 22 images at 129\,GHz. 
We excluded data from several epochs at certain frequencies when the observations clearly suffered from large pointing errors or bad weather.
The imaging process started with self-calibrating the visibility data based on a point source model of 1~Jy (i.e., the STARTMOD task in Difmap).
The CLEANing was performed until the residuals reached a certain total flux level, and then CLEAN was repeated with loops of phase self-calibration.
The standard CLEANing and self-calibration within the central emission regions have been conducted until no significant flux density was added compared to the image rms level.
Since OJ\,287 is very compact on mas-scales with the KVN observations, the best CLEAN models to the data can be found quickly.

\begin{figure*}
\centering
\includegraphics[width=0.9\textwidth]{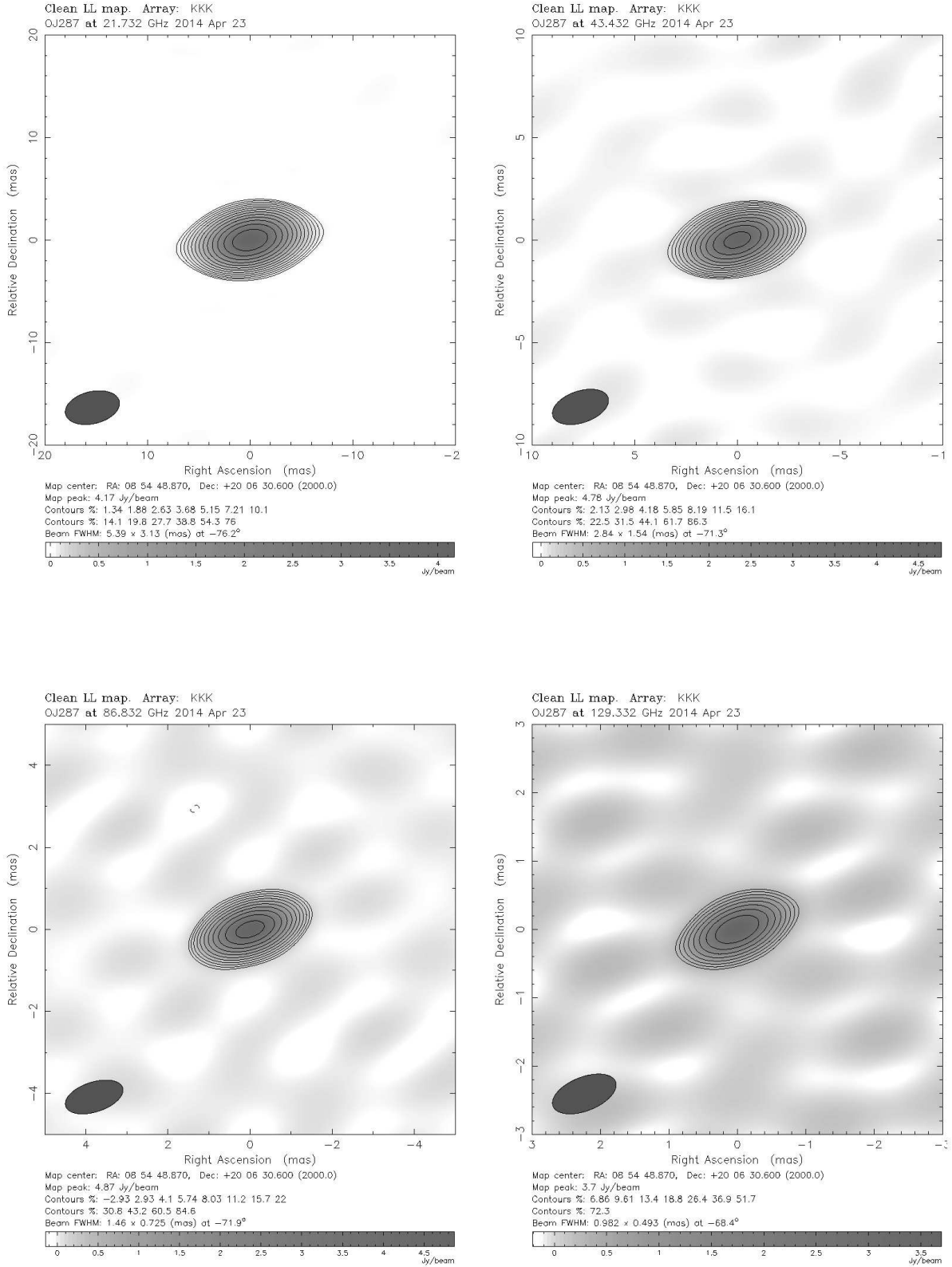}
\caption{\label{fig:clean_map}The CLEAN maps of OJ\,287 obtained in Epoch 14 (2014 April 23), showing compact core dominated structure at all frequencies from 22 to 129\,GHz. In each panel, the contours start from three times the rms noise level and increase by factors 2.
Left top indicates 22\,GHz, right top 43\,GHz, Left bottom 86\,GHz and right bottom 129\,GHz.}
\end{figure*}

Figure \ref{fig:clean_map} presents representative CLEAN contour maps of OJ\,287 observed at 22 (top left), 43 (top right), 86 (bottom left), and 129\,GHz (bottom right).
The source has a compact structure on mas-scales at all observing frequencies with the KVN. 
In order to evaluate the image quality, we computed the image quality factor. 
This is the ratio of the image root-mean-square noise and its mathematical expectation,  $\xi_{\rm r}=S_{\rm r}/S_{\rm r,exp}$, where $S_{\rm r}$ is the maximum absolute flux density in the residual image and $S_{\rm r,exp}$ is the expectation of $S_{\rm r}$. For more detailed descriptions of this image quality evaluation scheme (see, \citealt{Lobanov+06,Lee+16a,Lee+17a}).
$\xi_{\rm r}$ obtained in this paper is 0.42-0.89 which indicates images represent sufficiently the structure detected in the visibility data although the image model with $\xi_{\rm r}$ < 1 has a large number of degrees of freedom.

We conducted model-fitting procedures in order to parameterize physical parameters such as size, flux density, locations, etc. 
When the fitted component size is smaller than minimum resolvable size according to \cite{Lee+16a}, the component is considered to be unresolved. The minimum resolvable size was estimated for the core component to determine the upper limit following \cite{Lobanov+05}. The visibility data have been fitted with circular two-dimensional Gaussian components using the MODELFIT task in Difmap. If available, additional Gaussian components have been added until the $\chi^2$ value of the fitting is no longer improved.

The brightness temperature $T_{\rm b}$ was estimated as following $T_{b}=\frac{2ln2}{\pi k}\frac{S_{tot}\lambda^2}{d^2}(1+z)$, where $\lambda$ is the wavelength of the observation, $z$ is the redshift, d is the fitted core size, and $k$ is the Boltzmann constant.
Table \ref{table:image_para} lists the fitted parameters of the contour maps presented in Figure \ref{fig:clean_map}, including the observing frequencies, restoring beam size $B_{\rm maj,min} $, position angle $B_{\rm PA}$, total CLEANed flux density $S_{\rm KVN}$, peak flux density $S_{\rm p}$ in units of Jy per beam, rms in the image $\sigma$, the dynamic range (ratio of peak to rms) of the image $D$, and image quality factor $\xi_{\rm r}$.
The circular Gaussian model-fit parameters, total flux density $ S_{\rm tot}$, peak flux density $S_{\rm peak}$, size of the core component $d$, and brightness temperature $T_{\rm b}$ are given in Table \ref{table:modelfit}.

\subsection{OVRO 15\,GHz data}\label{subsec:ovro_data}
We collected the 15\,GHz light curve observed by the Owens Valley Radio Observatory (OVRO) 40-meter Telescope.
Since late 2007, The OVRO telescope has been operating monitoring observations of about 1500 blazars. 
The mean cadence of the monitoring is twice a week with a typical error on the flux density of 4\,mJy and systematic uncertainty of ~3\,\% \citep{Richards+11,Max+12}\footnote{http://www.astro.caltech.edu/ovroblazars}.
The data span we used in this paper is from 2013 January 1 (MJD 56301) to 2016 December 28 (MJD 57750).

\subsection{SMA 225\,GHz data}\label{subsec:sma_data}
To extend the multi-frequency data of the source to the higher frequencies, we collected Submillimeter Array (SMA) 1\,mm monitoring data.
The SMA is an interferometer composed of eight telescopes located near the summit of Mauna Kea, Hawaii. 
OJ\,287 is one of more than 400 sources that are monitored at 1.3\,mm and 850\,$\mu m$ wavelengths. 
Observations of available potential calibrators are from time to time observed for 3 to 5 minutes, and the measured source signal strength is calibrated by using known standards, typically solar system objects (Titan, Uranus, Neptune, or callisto).
More detailed explanations for the monitoring program of calibrators are described in \citet{Gurwell+07}.
Data from the program are updated regularly and are available at the SMA website\footnote{http://sma1.sma.hawaii.edu/callist.html}.
We exploited the 1\,mm SMA data of OJ\,287 from 2013 January 24 (MJD 56317) to 2016 October 13 (MJD 57513).

 \subsection{VLBA 43\, GHz data}\label{subsec:vlba}
OJ\,287 has been conducted with the VLBA observations as part of VLBA Boston University monthly monitoring program at 43\,GHz \citep{Jorstad+05}.
We used the data only to analysis of the magnetic field strengths in this works from 2014 January 20 (MJD 56677) to 2016 March 19 (MJD 57466).  In order to estimate the size and flux density of emission features within the map, we fitted circular Gaussians using the MODELFIT task in Difmap. 

\section{Results}\label{sec:result}

\subsection{VLBI morphology}\label{subsec:VLBI}
Typically a blazar jet exhibits a core-jet structure. The core is usually defined as the bright compact region that is observed at the most upstream point of the jet \citep{Jorstad+17}. Jet emission is observed to move out from the core. At high resolution OJ~287 is known to have a core and also a downstream quasi-stationary feature \citep{Agudo+12, Hodgson+16}. The source at all KVN observing frequencies is unresolved appearing point-like, with no jet emission and also the quasi-stationary feature unresolved (see, Figure \ref{fig:clean_map}). We therefore refer to bright unresolved region in the KVN maps as the core. 

\subsection{Multi-frequency light curves}\label{subsec:lc}
\begin{figure*}
\centering
\includegraphics[width=0.8\textwidth]{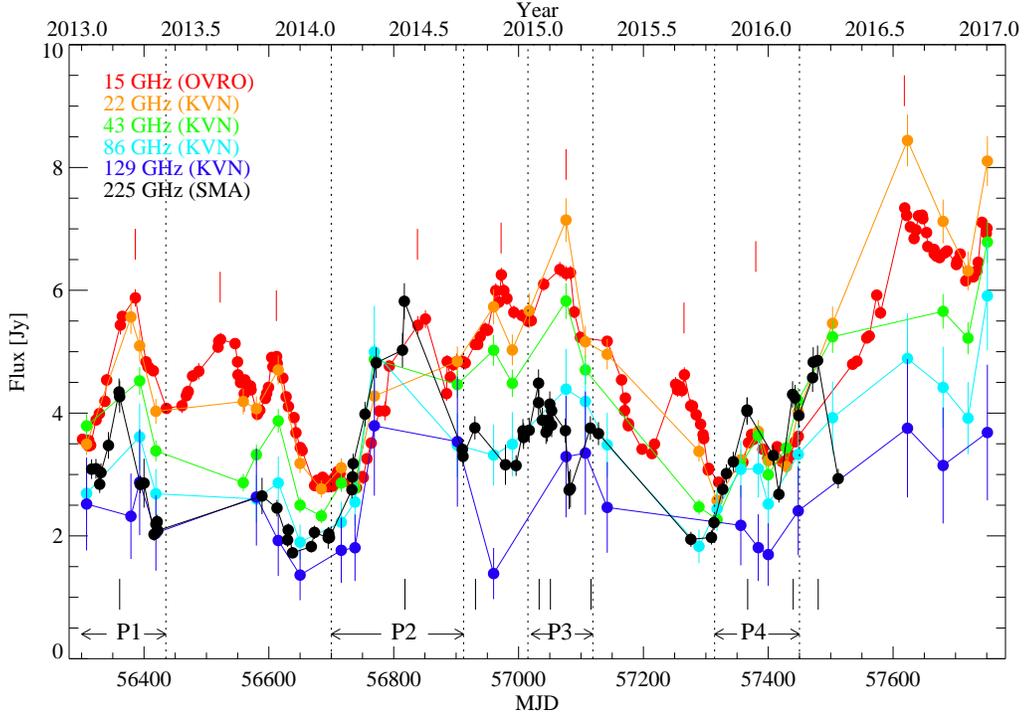}
\caption{\label{fig:lc_overlap}Light curves of OJ~287 from 15 to 225\,GHz observed from 2013 January until 2017 December. Red, yellow, green, skyblue, blue, and black symbols correspond to flux density at 15, 22, 43, 85, 129, 225\,GHz, respectively. The red and black bars indicate local peaks at 15 and 225\,GHz, respectively. Local peaks are labeled as P1, P2, P3, and P4, respectively, and are also marked dotted lines and horizontal arrows.}
\end{figure*}

We present the results and analysis of the flux density variability at multiple frequencies.
Figure \ref{fig:lc_overlap} shows multi-frequency light curves of OJ287 at 15\,GHz (OVRO), 22-129\,GHz (KVN), and 225\,GHz (SMA) observed between 2013 January 16 (MJD 56308) to 2016 December 28 (MJD 57750). The light curves show several flux enhancements over four years. 
We found 9 local peaks at 15\,GHz on MJD 56386, 56522, 56612, 56838, 56972, 57076, 57265, 57380, and 57618 and 9 local peaks at 225\,GHz on MJD 56361, 56818, 56931, 57033, 57051, 57116, 57367, 57440, and 57480. These local peaks are indicated by red vertical bars for 15\,GHz and black for 225\,GHz.  
We selected four parts (P1, P2, P3, and P4) which include local peaks in the KVN data also. These parts are marked within the dotted line sections as shown in Figure \ref{fig:lc_overlap}, in order to analyze the properties of the variability in flux density.

\subsubsection{The 15~GHz light curves}\label{subsec:ovro_lc}
The light curve at 15\,GHz with a mean cadence of 8\,days shows 9 local peaks over $\sim$ 4\,years in Figure \ref{fig:lc_overlap}. Over the period between MJD 56301 and 56673, three local peaks appeared and then the flux density was observed to be at a minimum of roughly 3\,Jy and which lasted $\sim$70\,days. 
The first local peak was designated as P1.
From MJD 56740 until 57320, four local peaks occurred and then the flux density was again observed to be at a minimum of roughly 3\,Jy. 
The first peak and the third peak of the five peaks were named as P2 and P3, respectively.
Between MJD 57320 and 57450, one small local peak (named P4) appeared on MJD 57380 and then the flux density significantly increased and local peak flux density of 7.3\,Jy was observed on MJD 57618.
In summary, the 15\,GHz light curve of OJ\,287 shows nine flux enhancements, ranging from roughly 3\,Jy to 7.3\,Jy with a mean flux density of 4.7\,Jy over four years.

\subsubsection{The 225~GHz light curves}\label{subsec:sma_lc}
The SMA 225\,GHz light curve is has periods of high cadence but also periods with sparse observations. Several of the flux enhancements observed at 15\,GHz were not traced but the trend of the light curve follows the 15\,GHz light curve quite well. 
Over the period between MJD 56301 and 56450 (P1), the first local peak appeared with flux density of 4.3\,Jy.
The light curve of 225\,GHz shows a major flux enhancement between MJD 56638 and MJD 56911 (P2) and subsequent flux enhancements appeared between MJD 57015 and MJD 57119 (P3), and MJD 57314 and MJD 57450 (P4).
The minimum and maximum flux density at 225\,GHz are 1.7\,Jy and 5.8\,Jy, respectively.

\subsubsection{The 22-129\,GHz light curves}\label{subsec:kvn_lc}
The light curves at 22-129 GHz were obtained simultaneously at an average monthly interval using the KVN. 
The number of available flux density measurements are 31 at 22\,GHz, 27 at 43\,GHz, 25 at 86\,GHz, and 19 at 129\,GHz.

Despite the relatively low cadence of the KVN observations, the trend of the KVN light curves is similar to the 15\,GHz and the 225\,GHz light curves (i.e., following increasing and decreasing trends, and showing the corresponding local minima of the flux density). 
The minimum and maximum flux densities are are 2.6\,Jy and 8.4\,Jy at 22\,GHz, 2.3\,Jy and 6.8\,Jy at 43\,GHz, 1.8\,Jy and 5.9\,Jy at 86\,GHz, 1.4\,Jy and 3.8\,Jy at 129\,GHz, respectively. The mean flux densities are 4.6\,Jy, 4.0\,Jy, 3.4\,Jy, and 2.5\,Jy at 22, 43, 86, and 129\,GHz, respectively.

\subsection{Fractional variability amplitude}\label{subsec:f_var}
There are several ways of measuring the amplitude of the flux density variability of blazar sources such as the variability index $(\rm V=(S_{max}-S_{min})/(S_{max}+S_{min}), S_{max}$ and $\rm S_{min}$ are maximum flux density and minimum flux density, respectively.) \citep{Aller+99}, the modulation index $(\rm m=100\cdot\sigma_{s}/<S>$, $\sigma_{s}$ and $<S>$ are rms of the flux density and mean flux density, respectively.) \citep{Kraus+03}, and the fractional variability amplitude \citep{Edelson+02}. We adopted the fractional variability amplitude method, which can additionally correct noise effects from the measurement errors \citep{Chidiac+16}.
The fractional variability amplitude, $F_{\rm var}$, is defined as 
\begin{equation}\label{eq:frac}
F_{\rm var}=\sqrt{\frac{V-\overline{\sigma^{2}_{err}}}{\bar{S}^2}},
\end{equation}
where $V$, $\overline{\sigma^2_{err}}$, and $\bar{S}$ are the variance of the flux density, mean squared error and mean flux density, respectively.
The uncertainty of the fractional variability is given by
\begin{equation}\label{eq:frac_err}
err(F_{\rm var})=\sqrt{\bigg(\sqrt{\frac{1}{2N}}\frac{\overline{\sigma^2_{err}}}{\bar{S}^2F_{var}}\bigg)^2+\bigg(\sqrt{\frac{\overline{\sigma}^2_{err}}{N}}\frac{1}{\bar{S}}\bigg)^2}.
\end{equation}

In Table \ref{tab:frac_var} we listed the estimated values of the fractional variability at 15~GHz to 225~GHz. 
For the 129\,GHz data, we could not calculate the fractional variability amplitude due to large measurement errors which is larger than the variance of the flux density.
Therefore, we exclude the 129 GHz data in the analysis.
The range of $F_{\rm var}$ was 0.25 to 0.33, which was relatively constant within the observing frequencies from 15 to 225\,GHz. This suggested that there was not significant variations in $F_{\rm var}$ in the frequency domain.
\begin{figure}
\centering
\includegraphics[width=0.5\textwidth]{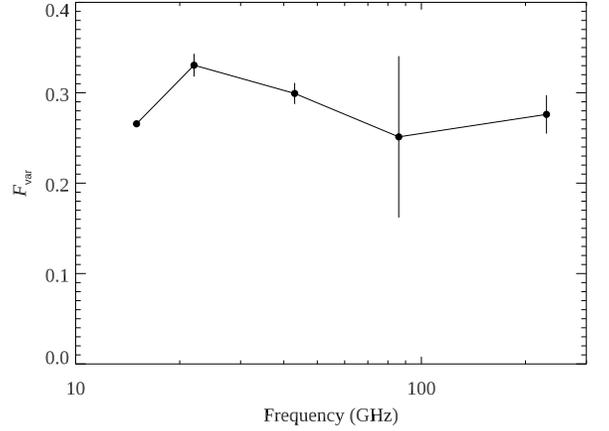}
\caption{\label{fig:frac_var}Fractional variability as a function of frequency of OJ~287.}
\end{figure}

\begin{table}
\centering
\caption{\label{tab:frac_var}Fractional variability}
\begin{tabular}{lr}
\hline
\hline
Frequency (GHz) & $F_{\rm var}$ \\
\hline
15 & 0.27$\pm$0.003\\
22 & 0.33$\pm$0.013\\
43 & 0.30$\pm$0.012\\
86 & 0.25$\pm$0.089\\
129 & 0.04$\pm$0.33\\
225 & 0.28$\pm$0.021\\
\hline
\end{tabular}
\end{table}

\subsection{Variability time scales at 15 and 225GHz}\label{sec:sf}
\begin{figure*}
\centering
\includegraphics[width=0.8\textwidth]{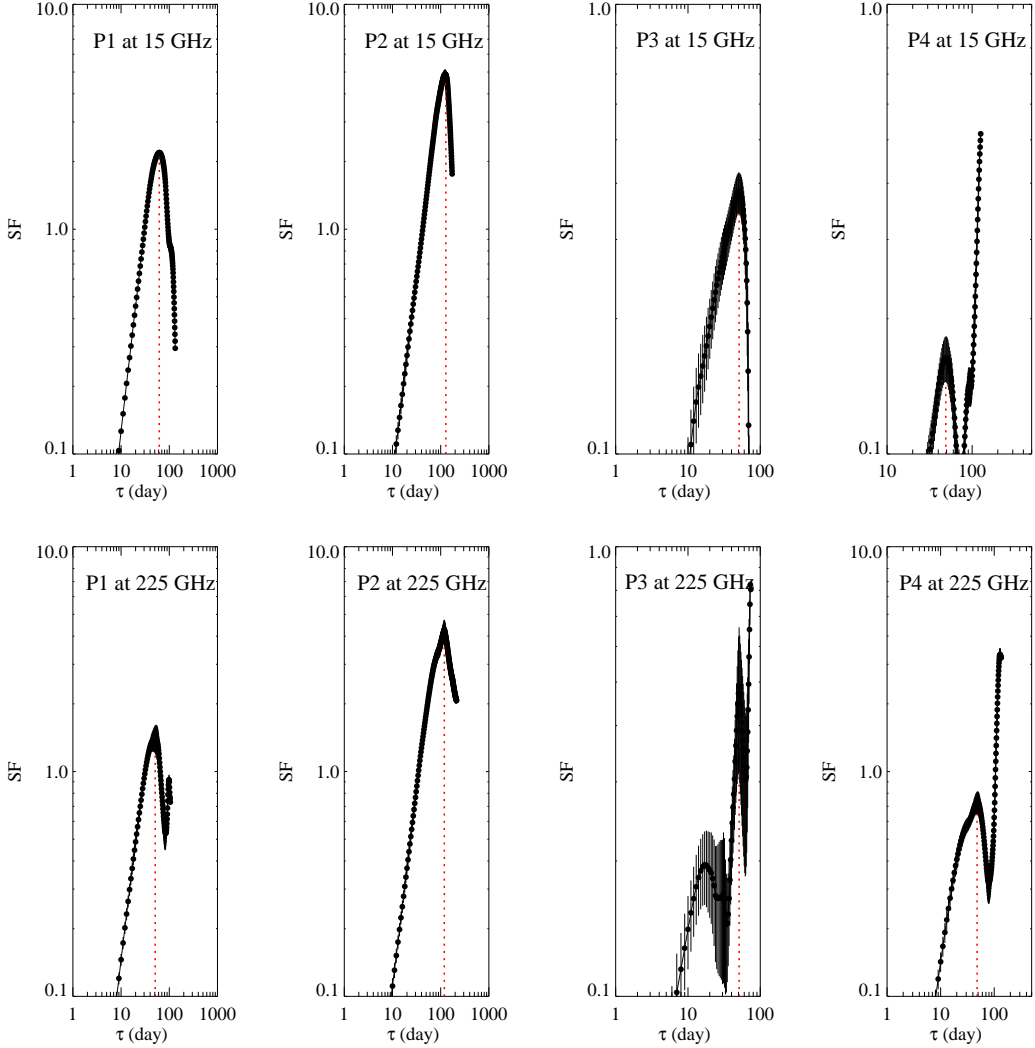}
\caption{\label{fig:sf}Structure function of OJ\,287 light curves at multiple radio frequencies. Top is P1-P4 at 15\,GHz, bottom is P1-P4 at 225\,GHz. Red dotted lines indicate time scales.}
\end{figure*}

In order to estimate the time scales of the individual flux enhancements, the light curves were linearly interpolated to 1-day intervals and then we employed the structure function (SF) to estimate the variability time scale as defined by \citet {Simonetti+85} and \citet{Heeschen+87}:
\begin{equation}\label{eq:sf}
SF(\tau)=\frac{1}{N}\sum_{i=1}^{N}[f(t_{i})-f(t_{i}+\tau)]^2,
\end{equation} 
where $f(t_{i})$, $\tau$, and $N$ are the flux density at time $t_{i}$, the time lag, and the number of data points, respectively. 
The SF curves for P1 - P4 at 15\,GHz and 230\,GHz are presented in Figure \ref{fig:sf}.
For the other frequencies (22-129\,GHz), the data are sparse and therefore we can not find reliable time scales for P1 to P4, individually.
The SF curves show a steeply rising trend to the first peak corresponding to characteristic time scale.
For P3 at 225\,GHz, the SF curve shows a double peak curve which indicates that there might be double peaks in the light curves. We used the second peak as it is considerably more significant. 

We obtained time scales $\tau_{\rm var}$ of 62$\pm$8\,days at 15\,GHz and 51$\pm$8\,days at 225\,GHz for P1, 128$\pm$12\,days at 15\,GHz and 119$\pm$21\,days at 225\,GHz for P2, 51$\pm$10\,days at 15\,GHz and 51$\pm$5\,days at 225\,GHz for P3, 49$\pm$8\,days at 15\,GHz and 48$\pm$12\,days at 225\,GHz for P4 of the time scales $\tau_{var}$ from the estimation of the SF as summarized in Table \ref{table:tb}.
Here the errors of the time scales indicate the median interval of light curves.
The corresponding SF values at the peak are 2.2, 4.9, 0.38, and 0.16 at 15\,GHz for P1-P4, respectively and 1.4, 4.3, 0.49, and 0.73 at 225\,GHz for P1-P4, respectively as marked with red dotted lines in Figure \ref{fig:sf}.
The estimated time scales of the three flux enhancements P1, P3, and P4 are similar with a time scale of $\sim$50 days at two frequencies, but the time scale of P2 is twice as long at the high frequency.
Since the SF value is proportional to the square of the amplitude of the flux enhancement, the strength of the flux enhancements for P1 and P2 is relatively larger in comparison with those of P3 and P4. 

\begin{table*}
\centering
\tabletypesize{\scriptsize}
\caption{\label{tab:tb_delta_size} Variability time sale, apparent brightness temperature, Doppler factor, size, time delay, and DCF}
\label{table:tb}
\begin{tabular}{ccccccccccc}
\hline
\hline
\multicolumn{8}{c}{P1} \\
\hline
$\nu$  & $\tau_{\rm var}$ & $\Delta S$ & $T_{\rm B,var}$ & $\delta_{var}$ & $\theta_{var}$ & $\tau$ & $\rm DCF$\\
(GHz)  &    (days)   &  (Jy) &   (K)         &                &   (mas) & (day) & \\
\hline
15 & 62$\pm$8 & 2.4$\pm$0.8 & 9.0$\times10^{14}$  & 7.8 & 0.067  & \multirow{2}{*}{-23$\pm$12.7} & \multirow{2}{*}{0.8} \\
225 & 51$\pm$8 & 2.3$\pm$0.7 & 5.4$\times10^{12}$ & 2.0 &  0.014 \\ 
\hline
\multicolumn{8}{c}{P2}\\
\hline
15  & 128$\pm$12 & 2.7$\pm$0.9 & 2.4$\times10^{14}$ & 5.5 & 0.098 & \multirow{2}{*}{-30.6$\pm$5.1} & \multirow{2}{*}{0.7}\\
225 & 119$\pm$21 & 3.8$\pm$1.2 & 1.6$\times10^{12}$ & 1.5 & 0.025 \\
\hline
\multicolumn{8}{c}{P3} \\
\hline
15  & 51$\pm$10 & 1.1$\pm$0.4 & 6.1$\times10^{14}$ & 7.1 & 0.050 & \multirow{2}{*}{-14.8$\pm$4.7} & \multirow{2}{*}{0.7}\\
225 & 51$\pm$5 & 1.7$\pm$0.5 & 4.0$\times10^{12}$ & 1.9 & 0.013 \\
\hline
\multicolumn{8}{c}{P4}\\
\hline
15  & 49$\pm$8 & 0.8$\pm$0.2 & 4.6$\times10^{14}$ & 6.6 & 0.045 & \multirow{2}{*}{-7.0$\pm$9.4} & \multirow{2}{*}{0.8}\\
225 & 48$\pm$12 & 2.1$\pm$0.7 & 5.5$\times10^{12}$ & 2.0 & 0.014 \\
\hline
\end{tabular}
\end{table*}

\section{Analysis}\label{sec:analysis}
\subsection{Spectral index}\label{subsec:sindex}

\begin{figure}
\centering
\includegraphics[width=0.5\textwidth]{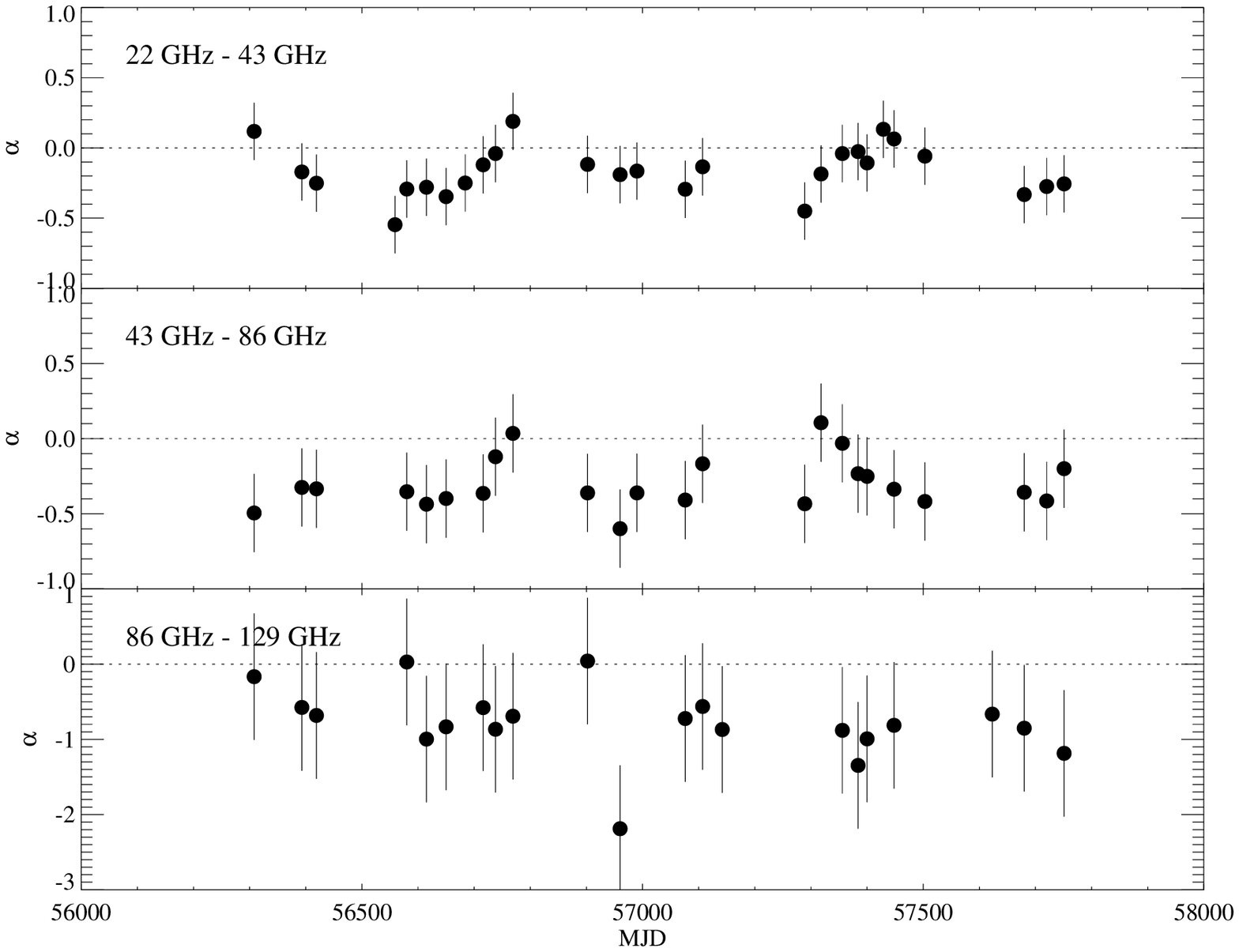}
\caption{\label{fig:sindex} Spectral indices of OJ 287. Top, middle, and bottom panels correspond to spectral indices estimated between 22 and 43\,GHz, 43 and 86\,GHz, and 86 and 129\,GHz, respectively.}
\end{figure}

The iMOGABA program facilitates spectral index studies, since the observations at the various frequencies within the program are exactly simultaneous.
Figure \ref{fig:sindex} shows the spectral index as a function of time for pairs of the KVN frequencies.
The spectral indices are variable as a function of time in the range of -$0.5\pm0.2$ to $0.2\pm0.2$ between 22 and 43\,GHz with mean spectral index, $<\alpha>_{22-43}$, of $-0.2$, $-0.6\pm0.3$ to $0.1\pm0.3$ between 43 and 86\,GHz with $<\alpha>_{43-86}$ of $-0.3$, $-2.2\pm0.8$ to $0.04\pm0.8$ between 86 and 129\,GHz with $<\alpha>_{86-129}$ of $-0.8$.
The spectral indices are mostly negative (i.e., optically thin) with a mean index as listed in Table \ref{table:sindex}.
The spectral index analysis implies that the source has an optically thin spectrum at millimeter wavelengths.

\begin{table}
\centering
\tabletypesize{\scriptsize}
\caption{\label{table:sindex}Mean spectral index}
\begin{tabular}{cc}
\hline
\hline
Frequency pair &  Mean spectral index\\
\hline
22-43\,GHz                     & -(0.15$\pm$0.20) \\
43-86\,GHz                    & -(0.32$\pm$0.26) \\
86-129\,GHz                  & -(0.79$\pm$0.84)\\
\hline
\end{tabular}
\end{table}

\subsection{Physical parameters from the variability time scales}

\subsubsection{Apparent brightness temperature}\label{subsubsec:tb}
Assuming that the component causing the flux variation has a spherical brightness distribution, its apparent brightness temperature can be estimated following \cite{Rani+13} as
\begin{equation}
T_{\rm B,var}=3.47\times10^5\times\Delta S\Bigg(\frac{\lambda D_{L}}{\tau_{var}(1+z)^2}\Bigg)^2,
\end{equation}

where $\Delta S$ is the variation of the flux density in Jy over a variability time scale, $\tau_{\rm var}$ in years, which was obtained from the structure function analysis in Section \ref{sec:sf}, $\lambda$ is the observing wavelength in cm, $D_{L}$ is the luminosity distance in Mpc, and $z$ is the redshift.
Here we adopt $D_{L}$=1630 \rm Mpc.
The values of $\tau_{var}$ and $\Delta S$ are listed in Table \ref{table:tb}.
We obtain apparent brightness temperatures $T_{\rm B,var}$ of $(2.4-9.0)\times10^{14}$\,K at 15\,GHz and $(1.6-5.5)\times10^{12}$\,K at 225\,GHz for P1-P4 as listed in Table~\ref{table:tb}.
We found that the $T_{\rm B,var,15}$ is higher than $T_{\rm B,var,225}$ in all flux enhancements.

\subsubsection{Doppler factor}\label{doppler_factor}

We obtained brightness temperature measurements at 15\,GHz and 225\,GHz in Section \ref{subsubsec:tb}. If we assume that there is a maximum intrinsic brightness temperature, $T_{\rm B,int}$, that can be achieved by the source we can estimate the Doppler factor of the source with the following expression\citep{Rani+13,Lee+17a} :

\begin{equation}
\delta_{var}=(1+z)\Bigg(\frac{T_{\rm B}^{\rm app}}{T_{\rm B,int}}\Bigg)^{1/(3+\alpha)},
\end{equation}
where $\alpha$ is the spectral index of optically thin regions. We adopted $\alpha$=$-0.8$ from the mean spectral index between 86 and 129\,GHz in our observations (see Section \ref{subsec:sindex}).
The Doppler factors estimated depend on the value chosen for $T_{\rm B,int}$. We have conservatively assumed a value of $10^{12}$ K, which is the inverse Compton limit \citep{Kellermann+69}. If the intrinsic brightness temperature limit is lower than $10^{12}$ K, this would increase the estimated values of the Doppler factors. 
The estimated Doppler factors are in the ranges of 5.5-7.8 at 15\,GHz and 1.5-2.0 at 225\,GHz for P1-P4 as listed in Table \ref{table:tb}.
The Doppler factors at 15\,GHz are higher than at 225\,GHz by a factor of at least $\sim$3. 
This implies that the emitting regions at two frequencies may be different.

\subsubsection{Size of the emitting region}\label{subsubsec:size}
Assuming that the variability of the flux density is intrinsic to the source, we can use causality arguments to estimate the size of the emitting region.
We can therefore compute the size of the emitting region, $\theta_{var}$ \cite{Rani+13}:

\begin{equation}
\theta_{var}=0.173\frac{\tau_{var}}{D_{L}}\delta_{var}(1+z),
\end{equation}

\noindent where $\tau_{var}$ is the time scale in days, $D_{L}$ is luminosity distance in Mpc and $\delta_{var}$ is the Doppler factor. 
The calculated sizes of emitting region are in the ranges of $0.058-0.128$\,mas at 15\,GHz and $0.017-0.032$\,mas at 225\,GHz for P1-P4 as tabulated in Table \ref{table:tb}.
The estimated sizes of emitting regions change according to the periods by a factor of 2 at 15\,GHz and 225\,GHz.
The mean sizes from the VLBA at 43\,GHz is 0.032$\pm$0.02 mas \citep{Jorstad+17} and from the GMVA at 86\,GHz is 0.035$\pm$0.02 mas \citep{Hodgson+17}.
The estimated $\theta_{var}$ compares well with the sizes obtained from these VLBI observations. 

\subsection{Discrete cross-correlation}\label{dcf}
Multi-frequency light curves enable us to study correlations with a time lag between different frequencies.
In order to compute the time lag between the flux density variability at 15 and 225\,GHz, we used discrete cross-correlation function (DCF) which is defined by \cite{Edelson+88}. 
A detailed description of the DCF can be found in \cite{Edelson+88} and \cite{Lee+17a}.
To calculate the DCF, we first compute unbinned discrete correlations ($\rm UDCF$) for two data sets as follows.

\begin{equation}
{\rm UDCF}_{ij}=\frac{(a_{i}-\bar{a})(b_{j}-\bar{b})}{\sqrt{\sigma^2_{a}-\sigma^2_{b}}},
\end{equation}

where $a_{i}$ and $b_{j}$ are the measurements of data sets $a$ and $b$
for each light curve, $\bar{a}$ and $\bar{b}$ are the mean of the data sets, and $\sigma_{a}$ and $\sigma_{b}$ are their corresponding standard deviations of the time series.
The $\rm UDCF$ is binned with an interval $\Delta\tau$
for estimating the $\rm DCF$
\begin{equation}
{\rm DCF}(\tau)=\frac{1}{M}\sum{\rm UDCF}_{ij}(\tau),
\end{equation}
for each time lag, where $M$ is the number of data points in the bin. 
We estimated the standard error for each bin with the following equation:
\begin{equation}
\sigma_{\rm DCF}(\tau)=\frac{1}{M-1}\sqrt{\sum[\rm UDCF_{ij}(\tau)-\rm DCF(\tau)]^2}.
\end{equation}
To calculate the DCF, the bin size has to be determined. We chose a minimum interval of 8 days in the light curve at 15\,GHz as the bin size. The correlated peak at a certain time in the DCF curves corresponds to the time delay between two light curves.
We performed a Monte Carlo simulation to estimate the uncertainty of the time lag obtained from the DCF.
We followed the random subset selection method as presented by \citet{Peterson+98} to estimate this uncertainty.

\begin{figure*}
\centering
\includegraphics[width=1.\textwidth]{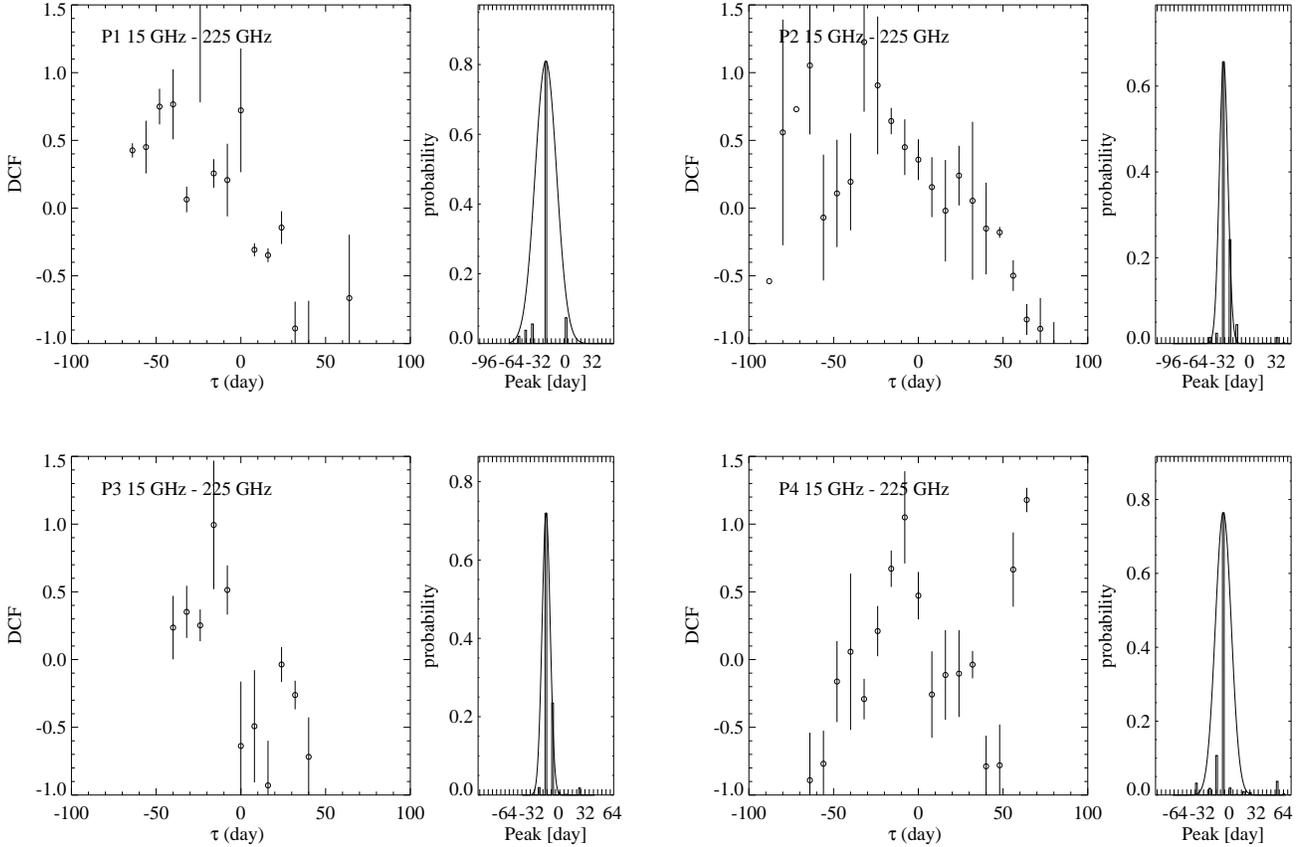}
\caption{\label{fig:dcf}Cross-correlation function and corresponding cross-correlation peak distribution of the light curves at 15 and 225 GHz in OJ~287 for P1-P4.}
\end{figure*}

From this method, randomly sampled light curves were generated $N$=1000 times to calculate the correlated peaks.
Therefore, we obtained 1000 time delays and their distribution, that is, cross-correlation peak distribution (CCPD) as shown in the right panel of Figure \ref{fig:dcf}.
Then we fitted the Gaussian function into the CCPD to measure the time delay.
We obtained the time delay of $-(23.0\pm12.7)$\,days for the period of P1 in the light curves, $-(30.6\pm5.1)$\,days for P2, $-(14.7\pm4.7)$\,days for P3, and $-(7.0\pm9.4)$\,days for P4.
If the CCPD can be assumed to be a normal distribution, a confidence interval of 95\,$\%$, $\tau_{95\,\%}$, can be calculated from $\tau - 1.96 \frac{\sigma}{\sqrt{N}} < \tau_{95\,\%} < \tau + 1.96 \frac{\sigma}{\sqrt{N}}$. Here $\tau$ is time delay, N is number of the simulations, and $\sigma$ is the standard deviation of the CCPD.
We found $-23.08 < \tau_{95\,\%} < -22.92$ for P1, $-30.68 < \tau_{95\,\%} < -30.52$ for P2, $-14.75 < \tau_{95\,\%} < -14.65$ for P3, and $-7.12 < \tau_{95\,\%} < -6.88$ for P4, respectively.
The minus sign means that the light curves at 225 GHz lead the one at 15 GHz as shown in the left panel of Figure. \ref{fig:dcf}. 
All the peaks of flux enhancements at 225\,GHz led those at 15\,GHz with a time lag of $7-30$ days.
The obtained time delay $\tau$ and the DCF values are listed in Table\,\ref{tab:tb_delta_size}.

\subsection{Radio spectral energy distribution}\label{subsec:nuc_sm}

The simultaneous multi-frequency observations at four frequencies (22, 43, 86, and 129\,GHz) enable us to study the millimeter-wavelength spectra of OJ\,287 without uncertainty in the time domain.

\begin{figure*}
\centering
\includegraphics[width=1.\textwidth]{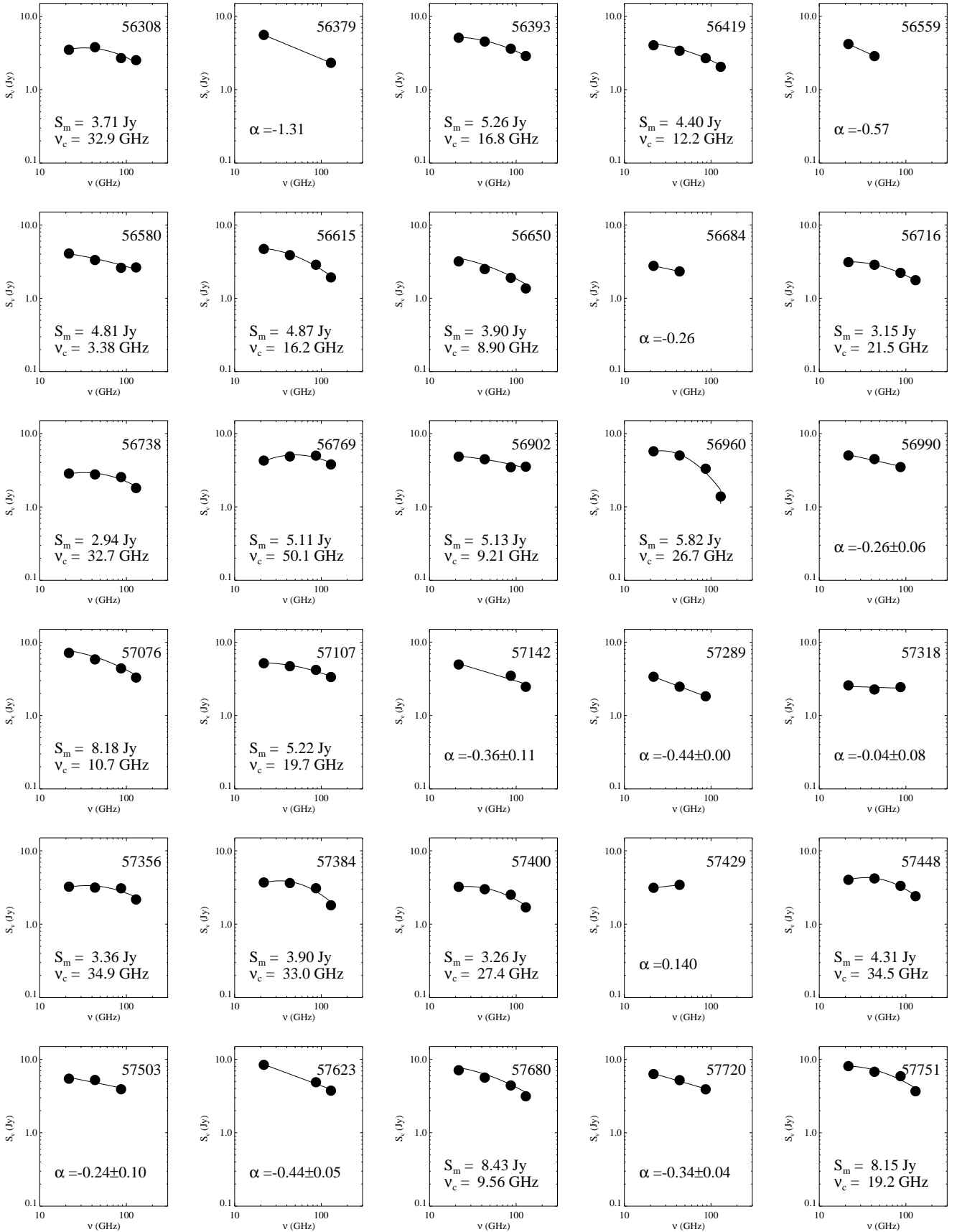}
\caption{\label{fig:curvefit}Spectral energy distribution of OJ~287 from 15 to 225 GHz observed over 2013 January to 2017 December. The 16th epoch (MJD 57017) is excluded since there was only one detection at 22\,GHz.}
\end{figure*}

The millimeter-wavelength spectrum has been obtained for those epochs when the source CLEAN flux density measurements are available at two to four frequency bands out of the 22, 43, 86, and 129~GHz bands. The spectrum with two or three measurements were fitted with a simple power-law function 

and that with four frequency measurements was fitted with a curved power-law function \citep{Lee+16a,Lee+17a,Algaba+18b} as
\begin{equation}
S=c_{1}\bigg(\frac{\nu}{\nu_{\rm c}}\bigg)^{a+c_{2}\rm ln(\nu/\nu_{\rm c})},
\end{equation}

\noindent where $S$ is the CLEAN flux density in Jy, $\nu$ is the observing frequency in GHz, $\nu_{c}$ is the turnover frequency, $a$ is the spectral index at $\nu_{c}$, and $c_{1}$ is a constant in Jy. Here $a$ can be considered to be the spectral index of optically thin emission.
In 11 epochs , there were less than 4 spectral data points. For these epochs, the spectrum was fitted with a power-law function. For the further 19 epochs, a curved power-law was used as it a produced a lower reduced $\chi^2$ value.  
For 8 epochs out of the curved power-law fit cases, we obtained turnover frequencies and peak flux densities within the observing frequency range of the KVN (see the panel (1) and (2) of Figure \ref{fig:lc_nuc_sm_bfield}).
The radio spectra of 30 epochs are plotted in Figure \ref{fig:curvefit} and the best-fit results are listed in Table \ref{table:nuc_sm_bfield}. 
In Figure \ref{fig:curvefit}, the spectral index $\alpha$ is obtained from the power-law fits and the turnover frequency $\nu_{c}$ and flux density $S_{m}$ at the turnover frequency are obtained from the curved power-law fits. 
All of the spectral indices $\alpha$ are steep with a minus sign in the range from -1.3 to -0.04 and the turnover frequency $\nu_{c}$ is in the range of 27 - 50\,GHz and the peak flux density $S_{m}$ is in the range of 2.9 - 5.8\,Jy, as summarized in Table\,\ref{table:nuc_sm_bfield}. 

\subsection{Synchrotron self-absorption magnetic field strength}\label{subsec:bssa}

The turnover frequency and peak flux density determined by the spectral model fitting (see Section \ref{subsec:nuc_sm}) allows us to constrain the magnetic field in the emission region that dominates the flux density at the turnover frequency, assuming that the emission region is a compact and hence is a homogeneous synchrotron self-absorption (SSA) region. Following \cite{Marscher+83} and \cite{Lee+17a}, we can express $B_{\rm SSA}$ as

\begin{equation}\label{eq:b_ssa}
B_{\rm SSA}=10^{-5}b(\alpha)S^{-2}_{\rm m}\theta^4\nu^5_{\rm c}\bigg(\frac{\delta}{1+z}\bigg)^{-1} (G),
\end{equation}

\noindent where $b(\alpha)$ is a parameter that depends on the optically thin spectral index ranging from 1.8 to 3.8 (see Table 1 in \cite{Marscher+83}, $\theta$ is 1.8 times the angular size (mas) of the emission region \citep{Marscher+77,Hodgson+16}. 
The observed size of SSA components are expected to vary as a function of frequency. This relationship under the model suggested by \citet{Blandford+79} is expected to follow the relationship $\theta \propto \nu^{-\epsilon}$. $\epsilon$ is a parameter related to the geometry of the jet. If $\epsilon$ =1, this would mean a fully conical jet and if $0 < \epsilon < 1$, this means the jet is parabolic. \cite{Algaba+17} found a value of $\epsilon=0.4\pm0.1$ for OJ\,287. We have adopted that value here. We linearly interpolated the VLBA sizes in time in order to estimate the size at the time of the measurements of the turnover frequency. We therefore used the following expression to estimate the size at the turnover frequency:
$\theta_{pred.}/\theta_{43}=(\nu_{c}/43)^{0.4\pm0.1}$.
The estimated sizes are listed in Table~\ref{table:nuc_sm_bfield}. 

The range of $\theta$ at the turnover frequency is 0.014-0.065\,mas.
We note that the Equation (\ref{eq:b_ssa}) is revised from the equation in \citet{Marscher+83} with the $(\delta/1+z)^{-1} $ factor instead of $(\delta/1+z)$ since the core of the source in the VLBI image map is thought to be quasi-stationary and therefore not moving.
The synchrotron self-absorption magnetic field strength was computed by performing N=1,000 Monte Carlo simulations via Eq.~(\ref{eq:b_ssa}), as described in detail in the following. 
All variables are simulated using a random sample of a normal distribution.
The simulated variables and errors indicate the mean values and standard deviation of normal distribution, respectively.
The input parameters are $\theta$, $\delta$, $\nu_{\rm c}$, and $S_{\rm m}$ and their respective errors.
Here, we used the values for $\theta$ showing in Table ~\ref{table:nuc_sm_bfield} and $\delta$=7.9 which are obtained from the VLBI observations by \cite{Hodgson+17}.
The value $b(\alpha)$ depends on the spectral index and its value can be found in \cite{Marscher+83}. The appropriate values have been listed in Table\,\ref{table:nuc_sm_bfield}. We used the turnover frequencies and flux density at the turnover frequency determined in Section~\ref{subsec:nuc_sm}.
The measured flux density at the turnover frequency $S_{\rm m}$=2.9-5.8 Jy, and the turnover frequency $\nu_{c}$=27.3-50.0 GHz, we obtained the magnetic field $B_{\rm SSA}$ ranging from 0.0004 to 0.255\,mG, as listed in Table~\ref{table:nuc_sm_bfield} and plotted panel (3) in Figure~\ref{fig:lc_nuc_sm_bfield}.

Due to the limited resolution of the KVN observations, special care must be taken to ensure that measured parameters are coming from either the core or the downstream quasi-stationary feature (see, Section~\ref{subsec:VLBI}). To check this, we plotted the flux density of the core and stationary feature in Figure \ref{fig:lc_bu} obtained from the VLBA-BU-BLAZAR monthly monitoring program at 43\,GHz. Only in epochs 12, 24, and 26 is the emission dominated by the core. The downstream quasi-stationary feature does not dominate the total flux density in any epochs. In these three epochs, the SSA magnetic field strength is measured to be roughly consistent at about 0.255 mG. This is approximately an order of magnitude lower than the magnetic field strength assuming equipartition.
In particular, in epoch 12, the turnover frequency was measured to be 50\,GHz which is close to the observing frequency of the BU program and also dominated by the core emission. 

\subsection{Equipartition Magnetic field strength}\label{subsec:beq}

\begin{table*}
\centering
\tabletypesize{\scriptsize}
\caption{Flux density at the turnover frequency $S_{\rm m}$, turnover frequency $\nu_{\rm c}$, angular size, synchrotron self-absorption magnetic field strength, and equipartition magnetic field strength. Only epochs 12, 24, and 26 are considered reliable for the magnetic field strength. See section \ref{subsec:bssa} for more details.} 
\label{table:nuc_sm_bfield}
\begin{tabular}{ccccccccc}
\hline
\hline
MJD   & Epoch & Date            & $S_{\rm m}$       & $\nu_{\rm c}$             & b$(\alpha)$ &$\theta$      & $B_{\rm SSA}$             & $B_{\rm eq}$  \\
(day) &              &                     &   (Jy)     &  (GHz)     &          & (mas)   &  (mG)      & (mG)           \\
\hline
56308 & 1     & 2013 Jan 16     & 3.7$\pm$0.1 & 32.7$\pm$3.3 & 3.1 & 0.032 & 0.025$\pm$0.025 & 1.82$\pm$0.28  \\
56738 & 11    & 2014 Mar 22  & 2.9$\pm$0.1 & 32.5$\pm$3.7 & 1.8 & 0.022 & 0.006$\pm$0.007 & 2.41$\pm$0.37  \\
56769 & 12    & 2014 Apr 23  & 5.1$\pm$0.1 & 50.0$\pm$2.0 & 3.5 & 0.036 & 0.161$\pm$0.146 & 1.93$\pm$0.30 \\
56960 & 14    & 2014 Oct 30   & 5.8$\pm$0.1 & 26.6$\pm$2.0 & 3.8 & 0.014 & 0.0004$\pm$0.0006 & 4.03$\pm$0.62  \\
57356 & 22    & 2015 Nov 30  & 3.4$\pm$0.1 & 34.7$\pm$3.7 & 1.7 & 0.063 & 0.226$\pm$0.144 & 1.02$\pm$0.16  \\
57384 & 23    & 2015 Dec 28  & 3.9$\pm$0.1 & 33.0$\pm$2.3 & 3.2 & 0.061 & 0.230$\pm$0.136& 1.07$\pm$0.17\\
57400 & 24    & 2016 Jan 13   & 3.3$\pm$0.1 & 27.3$\pm$3.7 & 3.2 & 0.065 & 0.157$\pm$0.104 & 0.95$\pm$0.15  \\
57448 & 26    & 2016 Mar 01 & 4.3$\pm$0.1 & 34.4$\pm$2.3 & 2.9 & 0.065 & 0.255$\pm$0.146 & 1.06$\pm$0.16 \\
\hline
\end{tabular}
\end{table*}

\begin{figure}
\centering
\includegraphics[width=0.5\textwidth]{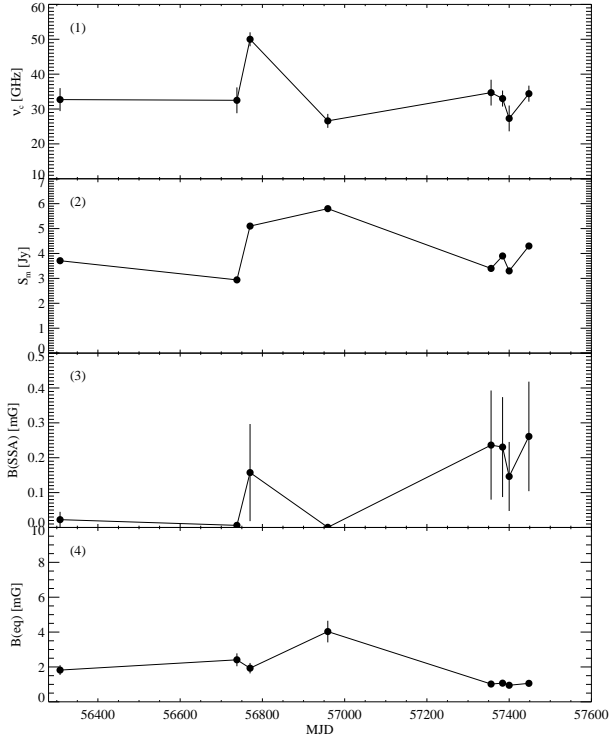}
\caption{\label{fig:lc_nuc_sm_bfield}(1) Turnover frequency $\nu_{c}$, (2) peak flux density $S_{m}$, (3) magnetic field strength derived from synchrotron self-absorption, and (4) equipartition magnetic field strength (4) of OJ~287.}
\end{figure}

\begin{figure}
\centering
\includegraphics[width=0.4\textwidth]{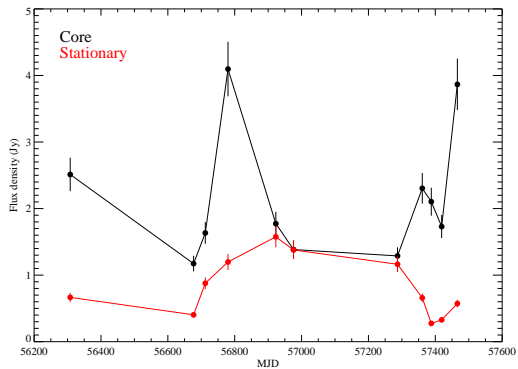}
\caption{\label{fig:lc_bu} Light curves of OJ~28. Black symbols indicate core and red symbols indicate stationary component that nearest to the core.}
\end{figure}

The magnetic field of the emission regions in the relativistic jet can also be investigated under the assumption that the energy densities of particles and magnetic fields are equal, following \cite{Kataoka+05} as

\begin{equation}
\begin{aligned}
 B_{\rm eq}=0.25\eta^{2/7}(1+z)^{11/7}{D_{\rm L}^{-2/7}}{\nu_{\rm c}}^{1/7} \\
 \times S_{\rm m}^{2/7}{\theta}^{-6/7}\delta^{-5/7}\quad \rm mG,
\end{aligned}
\end{equation}

where $\eta$ is the ratio of energy density carried by protons and electrons to the energy density of the electrons; i.e., the $\eta$= 1 for the leptonic jet (i.e., electron-positron) and $\eta$=1836 for the hadronic jet (i.e. electron-proton) \citep{Kataoka+05}. The parameters $D_{\rm L}$, $\nu_{\rm c}$, $S_{\rm m}$, $\theta_{\rm c}$, and $\delta$ are a luminosity distance in Mpc, turnover frequency in GHz, flux density at the turnover frequency in Jy, 1.8 times core size in mas, and Doppler factor, respectively.
For $\eta$=1, the magnetic field would be $\sim$3.8 times weaker and for $\eta$=1836, it would be $\sim$2.3 times stronger.
Here, we use $\eta=100$.
The values of $B_{\rm eq}$ are in the range of $0.95\pm0.15$ - $4.03\pm0.62$\,mG as listed in Table\,\ref{table:nuc_sm_bfield} and in panel (4) of Figure \ref{fig:lc_nuc_sm_bfield}.

\section{Discussion}\label{sec:discussion}
\subsection{Multi-frequency variability characteristics}
The calculated fractional variability amplitude in Section~\ref{subsec:f_var} shows no trend with frequency, with the exception of the 129\,GHz measurement which may be due to large uncertainties. The lack of a trend could be due to limited time sampling at higher frequencies. To test this possibility, we equalized the number of data points to that at 230\,GHz.  We then recalculated the fractional variability amplitude and found that there is no difference in the trend.
This constant trend is in contrast to the results of \citet{Algaba+18a} in 1633+382 and the results of \citet{Chidiac+16} in 3C~273. In both of these examples, increasing variability amplitudes were found as function of increasing frequency which was interpreted as being due to opacity effects. 
For individual flux enhancements (P1-P4), we performed the $F_{\rm var}$ test on each enhancement. We found that the results of the $F_{\rm var}$ test on individual flux enhancements are similar to that for whole light curves. We confirmed that the results of the $F_{\rm var}$ test on the radio/mm light curves of OJ287 are inconsistent with those for 1633+382 \citep{Algaba+18a} and 3C273 \citep{Chidiac+16}. 

This inconsistency may be due to the quasi-stationary component of OJ287. As shown in Figure 9, the trends of variability of the core and the quasi-stationary component are different from each other, resulting in the mixture variability of the integrated light curves obtained with single dish observations or low resolution VLBI observations at 15 to 225 GHz. For example, the core and the quasi-stationary component varies in flux density differently - the core and the quasi-stationary component gets brighter for P2 (MJD 56638 - 56910), whereas for P4 (MJD 57314 - 57450) the core gets brighter and the quasi-stationary component gets fainter. The mixture variability of multiple components may not clearly show the frequency dependence of the fractional variability.

\subsection{Time delay between 15 and 225\,GHz}

The source showed significant correlations in the flux density variability between the 15 and 225\,GHz light curves. The variability at the high frequency led that at 15\,GHz which can be interpreted by the opacity effect due to the synchrotron self-absorption of the spectrum. The source is optically thick at 15 GHz and optically thin at 225 GHz. Therefore, the opacity effect plays a role between these frequencies.
The obtained time delays between 15 and 225\,GHz light curves using the DCF analysis in Section \ref{dcf} are different for individual periods, i.e., $7-30$ days for Part 1 to Part 4 (over $\sim$4 years), the variation of the time delay, which was also reported for 3C~345 (a range of time delay of $0.4-1.8$ years between 5 and 37\,GHz over 18 years) by \citet{Kudryavtseva+11}.

The time delay $\Delta t$ can be described as 
$\Delta t=d \rm sin \theta/c\cdot\beta_{app}$
where $d$ is the distance between emitting regions in the jet, $\theta$ is the viewing angle of the jet to the line of sight, $c$ is the speed of light, and $\beta_{\rm app}$ is the apparent speed of the jet. \citet{Jorstad+17} reported the change of $\beta_{\rm app}$ in a range of $3.33-8.60$ and that of $\theta$ in a range of $1.3^{\circ} - 8.9^{\circ}$ over $\sim$5 years. 

These variations of those parameters may change the time delay by a factor of $<$ 8.8, assuming the distance between emitting regions is constant.
If we assume the $\beta_{\rm app}$ and $\theta$ are constant of 8.6 and 1.3${^\circ}$, respectively, the distance between 15 and 225\,GHz emitting regions can be estimated in a range of $2.2-9.7$\,pc.
Furthermore, if we assume the jet is conical at pc scale, we can derive the distance of the 225\,GHz emitting region to the jet apex using the size of the emitting region. 
Therefore the distance from the jet apex to 15\,GHz radio emitting regions are in the range of 3.2-12.9\,pc.
However, in \cite{Pushkarev+12}, the authors reported that the distance from the jet apex to the 15\,GHz core region was $<$4.13\,pc by using core-shift measurements assuming $\beta_{\rm app}$ of $\sim$15.2 and \citet{Hodgson+16} reported the distance from the jet apex to mm core is $<4-6$\,pc using magnetic filed relation of $B \propto r^{-1}$.
In some case, the distance we estimated is larger than the value they measured by a factor of $\sim$2-3. 
This difference may be explained by the changes of jet speed and viewing angle or the variability of the emitting region during the flaring or jet bending or core wandering \citep{Niinuma+15,Hodgson+17,Lisakov+17}.

\subsection{Magnetic field strengths}
After carefully analyzing the data in Section \ref{subsec:bssa} and \ref{subsec:beq}, we found only three epochs could be used to determined the magnetic field strengths. We found that within errors the $B_{\rm SSA}$ did not change. The $B_{\rm eq}$ is slightly lower in the last two epochs but given the assumptions and uncertainties we can not claim that this trend is significant. However, the difference between $B_{\rm SSA}$ and $B_{\rm eq}$ is an order of magnitude.
The assumptions in the equipartition measurements could account for an order of magnitude of difference \cite{Longair+94}. However, the difference here are large enough that we are confident in claiming that the source is not in equiparititon. That $B_{\rm SSA}$ is much smaller than $B_{\rm eq}$ suggests that the core is particle dominated, which is consistent with the results of \cite{Algaba+18b}. 

\section{Summary}\label{sec:summary}
We presented the results of VLBI observations of OJ~287 with the KVN at 22, 43, 86, and 129 GHz and combining the 15 GHz (OVRO) and 225 GHz (SMA) {\bf data over} $\sim$4 years from 2013 to 2017. We explored variability, physical parameters of the jet, and spectral variations by analyzing the multi-frequency light curves. We draw our conclusions as below.
\begin{enumerate}
\item OJ~287 shows compact core-dominated structure on the milliarcsecond scale at $22-129$\,GHz under the KVN resolutions of 1-6 mas.
\item The source shows several flux density enhancements at $15-225$\,GHz light curves which are divided into four parts designated as P1-P4. The time scales of the flux enhancements for P1-P4 are in the range of 49-128 days at 15\,GHz and of 48-119 days at 225\,GHz.
\item The fractional variability amplitudes in a range of $0.26-0.34$ are relatively constant over observing frequencies of $15-86$\,GHz.
\item The significant correlations between 15 and 225\,GHz light curves are found. The flux density variability at high frequency leads that at a low frequency which can be explained by the opacity effect.
\item The turnover frequency and peak flux density are in the range of $27-50$\,GHz and $3-6$\,Jy, respectively.
\item The synchrotron self-absorbed magnetic field strengths are in the range of $0.157\pm0.104-0.255\pm0.146$\,mG and the equipartition magnetic field strengths are in the range of $0.95\pm0.240.15-1.93\pm0.30$\,mG. The equipartition magnetic field strength is larger by a factor of 10 than synchrotron self-absorbed magnetic field, indicating that the downstream of the jet may be more particle dominated.
\end{enumerate}

\acknowledgments
We thank the anonymous referee for useful comments and suggestions that improved this paper.
We are grateful to all staff members in KVN
who helped to operate the array and to correlate the data.
The KVN is a facility operated by
the Korea Astronomy and Space Science Institute.
The KVN operations are supported
by KREONET (Korea Research Environment Open NETwork)
which is managed and operated
by KISTI (Korea Institute of Science and Technology Information).
J.-C.A acknowledges support from the Malaysian Fundamental Research Grant Scheme (FRGS) FRGS/1/2019/STG02/UM/02/6
This research has made use of data from the OVRO 40-m monitoring program \citep{Richards+11} which is supported in part by NASA grants NNX08AW31G, NNX11A043G, and 
NNX14AQ\\
89G and NSF grants AST-0808050 and AST-1109911.
The Submillimeter Array is a joint project between the Smithsonian Astrophysical Observatory and the Academia Sinica Institute of Astronomy and Astrophysics and is funded by the Smithsonian Institution and the Academia Sinica.
This study makes use of 43 GHz VLBA data from the VLBA-BU Blazar Monitoring Program (VLBA-BU-BLAZAR; http://www.bu.edu/blazars/VLBAproject.html), funded by NASA through the Fermi Guest Investigator Program. The VLBA is an instrument of the National Radio Astronomy Observatory. The National Radio Astronomy Observatory is a facility of the National Science Foundation operated by Associated Universities, Inc. 
SSL, SK, and SHK were supported by the National Research Foundation of Korea (NRF) grant funded by the Korea government (MIST) (2020R1A2C2009003).
MK were supported by JSPS KAKENHI Grant Numbers JP18K03656 and JP18H03721.

\bibliography{refer}

\begin{thebibliography}{}
\expandafter\ifx\csname natexlab\endcsname\relax\def\natexlab#1{#1}\fi
\providecommand{\url}[1]{\href{#1}{#1}}
\providecommand{\dodoi}[1]{doi:~\href{http://doi.org/#1}{\nolinkurl{#1}}}
\providecommand{\doeprint}[1]{\href{http://ascl.net/#1}{\nolinkurl{http://ascl.net/#1}}}
\providecommand{\doarXiv}[1]{\href{https://arxiv.org/abs/#1}{\nolinkurl{https://arxiv.org/abs/#1}}}

\bibitem[{{Agudo} {et~al.}(2012){Agudo}, {Marscher}, {Jorstad}, {G{\'o}mez},
  {Perucho}, {Piner}, {Rioja}, \& {Dodson}}]{Agudo+12}
{Agudo}, I., {Marscher}, A.~P., {Jorstad}, S.~G., {et~al.} 2012, \apj, 747, 63,
  \dodoi{10.1088/0004-637X/747/1/63}

\bibitem[{{Agudo} {et~al.}(2011){Agudo}, {Jorstad}, {Marscher}, {Larionov},
  {G{\'o}mez}, {L{\"a}hteenm{\"a}ki}, {Gurwell}, {Smith}, {Wiesemeyer}, {Thum},
  {Heidt}, {Blinov}, {D'Arcangelo}, {Hagen-Thorn}, {Morozova}, {Nieppola},
  {Roca-Sogorb}, {Schmidt}, {Taylor}, {Tornikoski}, \& {Troitsky}}]{Agudo+11}
{Agudo}, I., {Jorstad}, S.~G., {Marscher}, A.~P., {et~al.} 2011, \apjl, 726,
  L13, \dodoi{10.1088/2041-8205/726/1/L13}

\bibitem[{{Algaba} {et~al.}(2017){Algaba}, {Nakamura}, {Asada}, \&
  {Lee}}]{Algaba+17}
{Algaba}, J.~C., {Nakamura}, M., {Asada}, K., \& {Lee}, S.~S. 2017, \apj, 834,
  65, \dodoi{10.3847/1538-4357/834/1/65}

\bibitem[{{Algaba} {et~al.}(2015){Algaba}, {Zhao}, {Lee}, {Byun}, {Kang},
  {Kim}, {Kim}, {Kim}, {Kim}, {Kino}, {Miyazaki}, {Park}, {Trippe}, \&
  {Wajima}}]{Algaba+15}
{Algaba}, J.-C., {Zhao}, G.-Y., {Lee}, S.-S., {et~al.} 2015, Journal of Korean
  Astronomical Society, 48, 237, \dodoi{10.5303/JKAS.2015.48.5.237}

\bibitem[{{Algaba} {et~al.}(2018{\natexlab{a}}){Algaba}, {Lee}, {Kim}, {Rani},
  {Hodgson}, {Kino}, {Trippe}, {Park}, {Zhao}, {Byun}, {Gurwell}, {Kang},
  {Kim}, {Kim}, {Kim}, {Lott}, {Miyazaki}, \& {Wajima}}]{Algaba+18a}
{Algaba}, J.-C., {Lee}, S.-S., {Kim}, D.-W., {et~al.} 2018{\natexlab{a}}, \apj,
  852, 30, \dodoi{10.3847/1538-4357/aa9e50}

\bibitem[{{Algaba} {et~al.}(2018{\natexlab{b}}){Algaba}, {Lee}, {Rani}, {Kim},
  {Kino}, {Hodgson}, {Zhao}, {Byun}, {Gurwell}, {Kang}, {Kim}, {Kim}, {Kim},
  {Park}, {Trippe}, \& {Wajima}}]{Algaba+18b}
{Algaba}, J.-C., {Lee}, S.-S., {Rani}, B., {et~al.} 2018{\natexlab{b}}, \apj,
  859, 128, \dodoi{10.3847/1538-4357/aac2e7}

\bibitem[{{Aller}(1999)}]{Aller+99}
{Aller}, M.~F. 1999, in Astronomical Society of the Pacific Conference Series,
  Vol. 159, BL Lac Phenomenon, ed. L.~O. {Takalo} \& A.~{Sillanp{\"a}{\"a}}, 31

\bibitem[{{Angel} \& {Stockman}(1980)}]{Angel+80}
{Angel}, J.~R.~P., \& {Stockman}, H.~S. 1980, \araa, 18, 321,
  \dodoi{10.1146/annurev.aa.18.090180.001541}

\bibitem[{{Begelman} {et~al.}(1984){Begelman}, {Blandford}, \&
  {Rees}}]{Begelman+84}
{Begelman}, M.~C., {Blandford}, R.~D., \& {Rees}, M.~J. 1984, Reviews of Modern
  Physics, 56, 255, \dodoi{10.1103/RevModPhys.56.255}

\bibitem[{{Blandford} \& {K{\"o}nigl}(1979)}]{Blandford+79}
{Blandford}, R.~D., \& {K{\"o}nigl}, A. 1979, \apj, 232, 34,
  \dodoi{10.1086/157262}

\bibitem[{{Britzen} {et~al.}(2018){Britzen}, {Fendt}, {Witzel}, {Qian},
  {Pashchenko}, {Kurtanidze}, {Zajacek}, {Martinez}, {Karas}, {Aller}, {Aller},
  {Eckart}, {Nilsson}, {Ar{\'e}valo}, {Cuadra}, {Subroweit}, \&
  {Witzel}}]{Britzen+18}
{Britzen}, S., {Fendt}, C., {Witzel}, G., {et~al.} 2018, \mnras, 478, 3199,
  \dodoi{10.1093/mnras/sty1026}

\bibitem[{{Chidiac} {et~al.}(2016){Chidiac}, {Rani}, {Krichbaum}, {Angelakis},
  {Fuhrmann}, {Nestoras}, {Zensus}, {Sievers}, {Ungerechts}, {Itoh},
  {Fukazawa}, {Uemura}, {Sasada}, {Gurwell}, \& {Fedorova}}]{Chidiac+16}
{Chidiac}, C., {Rani}, B., {Krichbaum}, T.~P., {et~al.} 2016, \aap, 590, A61,
  \dodoi{10.1051/0004-6361/201628347}

\bibitem[{{Deller} {et~al.}(2007){Deller}, {Tingay}, {Bailes}, \&
  {West}}]{Deller+07}
{Deller}, A.~T., {Tingay}, S.~J., {Bailes}, M., \& {West}, C. 2007, \pasp, 119,
  318, \dodoi{10.1086/513572}

\bibitem[{{Deller} {et~al.}(2011){Deller}, {Brisken}, {Phillips}, {Morgan},
  {Alef}, {Cappallo}, {Middelberg}, {Romney}, {Rottmann}, {Tingay}, \&
  {Wayth}}]{Deller+11}
{Deller}, A.~T., {Brisken}, W.~F., {Phillips}, C.~J., {et~al.} 2011, \pasp,
  123, 275, \dodoi{10.1086/658907}

\bibitem[{{Edelson} {et~al.}(2002){Edelson}, {Turner}, {Pounds}, {Vaughan},
  {Markowitz}, {Marshall}, {Dobbie}, \& {Warwick}}]{Edelson+02}
{Edelson}, R., {Turner}, T.~J., {Pounds}, K., {et~al.} 2002, \apj, 568, 610,
  \dodoi{10.1086/323779}

\bibitem[{{Edelson} \& {Krolik}(1988)}]{Edelson+88}
{Edelson}, R.~A., \& {Krolik}, J.~H. 1988, \apj, 333, 646,
  \dodoi{10.1086/166773}

\bibitem[{{Gurwell} {et~al.}(2007){Gurwell}, {Peck}, {Hostler}, {Darrah}, \&
  {Katz}}]{Gurwell+07}
{Gurwell}, M.~A., {Peck}, A.~B., {Hostler}, S.~R., {Darrah}, M.~R., \& {Katz},
  C.~A. 2007, in Astronomical Society of the Pacific Conference Series, Vol.
  375, From Z-Machines to ALMA: (Sub)Millimeter Spectroscopy of Galaxies, ed.
  A.~J. {Baker}, J.~{Glenn}, A.~I. {Harris}, J.~G. {Mangum}, \& M.~S. {Yun},
  234

\bibitem[{{Heeschen} {et~al.}(1987){Heeschen}, {Krichbaum}, {Schalinski}, \&
  {Witzel}}]{Heeschen+87}
{Heeschen}, D.~S., {Krichbaum}, T., {Schalinski}, C.~J., \& {Witzel}, A. 1987,
  \aj, 94, 1493, \dodoi{10.1086/114583}

\bibitem[{{Hodgson} {et~al.}(2016){Hodgson}, {Lee}, {Zhao}, {Algaba}, {Yun},
  {Jung}, \& {Byun}}]{Hodgson+16}
{Hodgson}, J.~A., {Lee}, S.-S., {Zhao}, G.-Y., {et~al.} 2016, Journal of Korean
  Astronomical Society, 49, 137, \dodoi{10.5303/JKAS.2016.49.4.137}

\bibitem[{{Hodgson} {et~al.}(2017){Hodgson}, {Krichbaum}, {Marscher},
  {Jorstad}, {Rani}, {Marti-Vidal}, {Bach}, {Sanchez}, {Bremer}, {Lindqvist},
  {Uunila}, {Kallunki}, {Vicente}, {Fuhrmann}, {Angelakis}, {Karamanavis},
  {Myserlis}, {Nestoras}, {Chidiac}, {Sievers}, {Gurwell}, \&
  {Zensus}}]{Hodgson+17}
{Hodgson}, J.~A., {Krichbaum}, T.~P., {Marscher}, A.~P., {et~al.} 2017, \aap,
  597, A80, \dodoi{10.1051/0004-6361/201526727}

\bibitem[{{Jorstad} {et~al.}(2005){Jorstad}, {Marscher}, {Lister}, {Stirling},
  {Cawthorne}, {Gear}, {G{\'o}mez}, {Stevens}, {Smith}, {Forster}, \&
  {Robson}}]{Jorstad+05}
{Jorstad}, S.~G., {Marscher}, A.~P., {Lister}, M.~L., {et~al.} 2005, \aj, 130,
  1418, \dodoi{10.1086/444593}

\bibitem[{{Jorstad} {et~al.}(2017){Jorstad}, {Marscher}, {Morozova},
  {Troitsky}, {Agudo}, {Casadio}, {Foord}, {G{\'o}mez}, {MacDonald}, {Molina},
  {L{\"a}hteenm{\"a}ki}, {Tammi}, \& {Tornikoski}}]{Jorstad+17}
{Jorstad}, S.~G., {Marscher}, A.~P., {Morozova}, D.~A., {et~al.} 2017, \apj,
  846, 98, \dodoi{10.3847/1538-4357/aa8407}

\bibitem[{{Kataoka} \& {Stawarz}(2005)}]{Kataoka+05}
{Kataoka}, J., \& {Stawarz}, {\L}. 2005, \apj, 622, 797, \dodoi{10.1086/428083}

\bibitem[{{Kellermann} \& {Pauliny-Toth}(1969)}]{Kellermann+69}
{Kellermann}, K.~I., \& {Pauliny-Toth}, I.~I.~K. 1969, \apjl, 155, L71,
  \dodoi{10.1086/180305}

\bibitem[{{Kim} {et~al.}(2018){Kim}, {Trippe}, {Lee}, {Kim}, {Algaba},
  {Hodgson}, {Park}, {Kino}, {Zhao}, {Wajima}, {Lee}, \& {Kang}}]{Kim+18}
{Kim}, D.-W., {Trippe}, S., {Lee}, S.-S., {et~al.} 2018, \mnras, 480, 2324,
  \dodoi{10.1093/mnras/sty1993}

\bibitem[{{Kraus} {et~al.}(2003){Kraus}, {Krichbaum}, {Wegner}, {Witzel},
  {Cim{\`o}}, {Quirrenbach}, {Britzen}, {Fuhrmann}, {Lobanov}, {Naundorf},
  {Otterbein}, {Peng}, {Risse}, {Ros}, \& {Zensus}}]{Kraus+03}
{Kraus}, A., {Krichbaum}, T.~P., {Wegner}, R., {et~al.} 2003, \aap, 401, 161,
  \dodoi{10.1051/0004-6361:20030118}

\bibitem[{{Kudryavtseva} {et~al.}(2011){Kudryavtseva}, {Gabuzda}, {Aller}, \&
  {Aller}}]{Kudryavtseva+11}
{Kudryavtseva}, N.~A., {Gabuzda}, D.~C., {Aller}, M.~F., \& {Aller}, H.~D.
  2011, \mnras, 415, 1631, \dodoi{10.1111/j.1365-2966.2011.18808.x}

\bibitem[{{Lee} {et~al.}(2017{\natexlab{a}}){Lee}, {Lee}, {Hodgson}, {Kim},
  {Algaba}, {Kang}, {Kang}, \& {Kim}}]{Lee+17a}
{Lee}, J.~W., {Lee}, S.-S., {Hodgson}, J.~A., {et~al.} 2017{\natexlab{a}},
  \apj, 841, 119, \dodoi{10.3847/1538-4357/aa72f7}

\bibitem[{{Lee} {et~al.}(2016{\natexlab{a}}){Lee}, {Lee}, {Kang}, {Byun}, \&
  {Kim}}]{Lee+16b}
{Lee}, J.~W., {Lee}, S.-S., {Kang}, S., {Byun}, D.-Y., \& {Kim}, S.~S.
  2016{\natexlab{a}}, \aap, 592, L10, \dodoi{10.1051/0004-6361/201629212}

\bibitem[{{Lee} {et~al.}(2017{\natexlab{b}}){Lee}, {Sohn}, {Byun}, {Lee}, \&
  {Kim}}]{Lee+17b}
{Lee}, J.~W., {Sohn}, B.~W., {Byun}, D.-Y., {Lee}, J.~A., \& {Kim}, S.~S.
  2017{\natexlab{b}}, \aap, 601, A12, \dodoi{10.1051/0004-6361/201629346}

\bibitem[{{Lee} {et~al.}(2013){Lee}, {Han}, {Kang}, {Seen}, {Byun}, {Baek},
  {Kim}, \& {Kim}}]{Lee+13}
{Lee}, S.-S., {Han}, M., {Kang}, S., {et~al.} 2013, in European Physical
  Journal Web of Conferences, Vol.~61, European Physical Journal Web of
  Conferences, 07007, \dodoi{10.1051/epjconf/20136107007}

\bibitem[{{Lee} {et~al.}(2015){Lee}, {Byun}, {Oh}, {Kim}, {Kim}, {Jung}, {Oh},
  {Roh}, {Jung}, \& {Yeom}}]{Lee+15}
{Lee}, S.-S., {Byun}, D.-Y., {Oh}, C.~S., {et~al.} 2015, Journal of Korean
  Astronomical Society, 48, 229, \dodoi{10.5303/JKAS.2015.48.5.229}

\bibitem[{{Lee} {et~al.}(2016{\natexlab{b}}){Lee}, {Wajima}, {Algaba}, {Zhao},
  {Hodgson}, {Kim}, {Park}, {Kim}, {Miyazaki}, {Byun}, {Kang}, {Kim}, {Kim},
  {Kino}, \& {Trippe}}]{Lee+16a}
{Lee}, S.-S., {Wajima}, K., {Algaba}, J.-C., {et~al.} 2016{\natexlab{b}},
  \apjs, 227, 8, \dodoi{10.3847/0067-0049/227/1/8}

\bibitem[{{Lisakov} {et~al.}(2017){Lisakov}, {Kovalev}, {Savolainen},
  {Hovatta}, \& {Kutkin}}]{Lisakov+17}
{Lisakov}, M.~M., {Kovalev}, Y.~Y., {Savolainen}, T., {Hovatta}, T., \&
  {Kutkin}, A.~M. 2017, \mnras, 468, 4478, \dodoi{10.1093/mnras/stx710}

\bibitem[{{Lobanov}(2005)}]{Lobanov+05}
{Lobanov}, A.~P. 2005, arXiv e-prints, astro.
\newblock \doarXiv{astro-ph/0503225}

\bibitem[{{Lobanov} {et~al.}(2006){Lobanov}, {Krichbaum}, {Witzel}, \&
  {Zensus}}]{Lobanov+06}
{Lobanov}, A.~P., {Krichbaum}, T.~P., {Witzel}, A., \& {Zensus}, J.~A. 2006,
  \pasj, 58, 253, \dodoi{10.1093/pasj/58.2.253}

\bibitem[{{Longair}(1994)}]{Longair+94}
{Longair}, M.~S. 1994, {High energy astrophysics. Volume 2. Stars, the Galaxy
  and the interstellar medium.}, Vol.~2

\bibitem[{{Marscher}(1977)}]{Marscher+77}
{Marscher}, A.~P. 1977, \apj, 216, 244, \dodoi{10.1086/155467}

\bibitem[{{Marscher}(1983)}]{Marscher+83}
---. 1983, \apj, 264, 296, \dodoi{10.1086/160597}

\bibitem[{{Max-Moerbeck} {et~al.}(2012){Max-Moerbeck}, {Richards}, {Pavlidou},
  {Hovatta}, {King}, {Pearson}, {Readhead}, {Reeves}, \& {Shepherd}}]{Max+12}
{Max-Moerbeck}, W., {Richards}, J.~L., {Pavlidou}, V., {et~al.} 2012, arXiv
  e-prints, arXiv:1205.0276.
\newblock \doarXiv{1205.0276}

\bibitem[{{Myserlis} {et~al.}(2018){Myserlis}, {Komossa}, {Angelakis},
  {G{\'o}mez}, {Karamanavis}, {Krichbaum}, {Bach}, \& {Grupe}}]{Myserlis+18}
{Myserlis}, I., {Komossa}, S., {Angelakis}, E., {et~al.} 2018, \aap, 619, A88,
  \dodoi{10.1051/0004-6361/201732273}

\bibitem[{{Niinuma} {et~al.}(2015){Niinuma}, {Kino}, {Doi}, {Hada}, {Nagai}, \&
  {Koyama}}]{Niinuma+15}
{Niinuma}, K., {Kino}, M., {Doi}, A., {et~al.} 2015, \apjl, 807, L14,
  \dodoi{10.1088/2041-8205/807/1/L14}

\bibitem[{{Nilsson} {et~al.}(2010){Nilsson}, {Takalo}, {Lehto}, \&
  {Sillanp{\"a}{\"a}}}]{Nilsson+10}
{Nilsson}, K., {Takalo}, L.~O., {Lehto}, H.~J., \& {Sillanp{\"a}{\"a}}, A.
  2010, \aap, 516, A60, \dodoi{10.1051/0004-6361/201014198}

\bibitem[{{Park} {et~al.}(2019){Park}, {Lee}, {Kim}, {Hodgson}, {Trippe},
  {Kim}, {Algaba}, {Kino}, {Zhao}, {Lee}, \& {Gurwell}}]{Park+19}
{Park}, J., {Lee}, S.-S., {Kim}, J.-Y., {et~al.} 2019, \apj, 877, 106,
  \dodoi{10.3847/1538-4357/ab1b27}

\bibitem[{{Peterson} {et~al.}(1998){Peterson}, {Wanders}, {Horne}, {Collier},
  {Alexander}, {Kaspi}, \& {Maoz}}]{Peterson+98}
{Peterson}, B.~M., {Wanders}, I., {Horne}, K., {et~al.} 1998, \pasp, 110, 660,
  \dodoi{10.1086/316177}

\bibitem[{{Pushkarev} {et~al.}(2012){Pushkarev}, {Hovatta}, {Kovalev},
  {Lister}, {Lobanov}, {Savolainen}, \& {Zensus}}]{Pushkarev+12}
{Pushkarev}, A.~B., {Hovatta}, T., {Kovalev}, Y.~Y., {et~al.} 2012, \aap, 545,
  A113, \dodoi{10.1051/0004-6361/201219173}

\bibitem[{{Rani} {et~al.}(2013){Rani}, {Krichbaum}, {Fuhrmann}, {B{\"o}ttcher},
  {Lott}, {Aller}, {Aller}, {Angelakis}, {Bach}, {Bastieri}, {Falcone},
  {Fukazawa}, {Gabanyi}, {Gupta}, {Gurwell}, {Itoh}, {Kawabata}, {Krips},
  {L{\"a}hteenm{\"a}ki}, {Liu}, {Marchili}, {Max-Moerbeck}, {Nestoras},
  {Nieppola}, {Quintana-Lacaci}, {Readhead}, {Richards}, {Sasada}, {Sievers},
  {Sokolovsky}, {Stroh}, {Tammi}, {Tornikoski}, {Uemura}, {Ungerechts},
  {Urano}, \& {Zensus}}]{Rani+13}
{Rani}, B., {Krichbaum}, T.~P., {Fuhrmann}, L., {et~al.} 2013, \aap, 552, A11,
  \dodoi{10.1051/0004-6361/201321058}

\bibitem[{{Richards} {et~al.}(2011){Richards}, {Max-Moerbeck}, {Pavlidou},
  {Readhead}, {Pearson}, {King}, {Reeves}, {Stevenson}, \&
  {Shepherd}}]{Richards+11}
{Richards}, J.~L., {Max-Moerbeck}, W., {Pavlidou}, V., {et~al.} 2011, arXiv
  e-prints, arXiv:1111.0318.
\newblock \doarXiv{1111.0318}

\bibitem[{{Rybicki} \& {Lightman}(1986)}]{Rybicki+86}
{Rybicki}, G.~B., \& {Lightman}, A.~P. 1986, {Radiative Processes in
  Astrophysics}

\bibitem[{{Sillanpaa} {et~al.}(1988){Sillanpaa}, {Haarala}, {Valtonen},
  {Sundelius}, \& {Byrd}}]{Sillanpaa+88}
{Sillanpaa}, A., {Haarala}, S., {Valtonen}, M.~J., {Sundelius}, B., \& {Byrd},
  G.~G. 1988, \apj, 325, 628, \dodoi{10.1086/166033}

\bibitem[{{Simonetti} {et~al.}(1985){Simonetti}, {Cordes}, \&
  {Heeschen}}]{Simonetti+85}
{Simonetti}, J.~H., {Cordes}, J.~M., \& {Heeschen}, D.~S. 1985, \apj, 296, 46,
  \dodoi{10.1086/163418}

\bibitem[{{Sitko} \& {Junkkarinen}(1985)}]{Sitko+85}
{Sitko}, M.~L., \& {Junkkarinen}, V.~T. 1985, \pasp, 97, 1158,
  \dodoi{10.1086/131679}

\bibitem[{{Spergel} {et~al.}(2015){Spergel}, {Flauger}, \&
  {Hlo{\v{z}}ek}}]{Spergel+15}
{Spergel}, D.~N., {Flauger}, R., \& {Hlo{\v{z}}ek}, R. 2015, \prd, 91, 023518,
  \dodoi{10.1103/PhysRevD.91.023518}

\bibitem[{{Urry} \& {Padovani}(1995)}]{Urry+95}
{Urry}, C.~M., \& {Padovani}, P. 1995, \pasp, 107, 803, \dodoi{10.1086/133630}

\bibitem[{{Valtonen} \& {Pihajoki}(2013)}]{Valtonen+13}
{Valtonen}, M., \& {Pihajoki}, P. 2013, \aap, 557, A28,
  \dodoi{10.1051/0004-6361/201321754}

\bibitem[{{Valtonen} {et~al.}(2006){Valtonen}, {Lehto}, {Sillanp{\"a}{\"a}},
  {Nilsson}, {Mikkola}, {Hudec}, {Basta}, {Ter{\"a}sranta}, {Haque}, \&
  {Rampadarath}}]{Valtonen+06}
{Valtonen}, M.~J., {Lehto}, H.~J., {Sillanp{\"a}{\"a}}, A., {et~al.} 2006,
  \apj, 646, 36, \dodoi{10.1086/504884}

\bibitem[{{Valtonen} {et~al.}(2008){Valtonen}, {Lehto}, {Nilsson}, {Heidt},
  {Takalo}, {Sillanp{\"a}{\"a}}, {Villforth}, {Kidger}, {Poyner}, {Pursimo},
  {Zola}, {Wu}, {Zhou}, {Sadakane}, {Drozdz}, {Koziel}, {Marchev}, {Ogloza},
  {Porowski}, {Siwak}, {Stachowski}, {Winiarski}, {Hentunen}, {Nissinen},
  {Liakos}, \& {Dogru}}]{Valtonen+08}
{Valtonen}, M.~J., {Lehto}, H.~J., {Nilsson}, K., {et~al.} 2008, \nat, 452,
  851, \dodoi{10.1038/nature06896}

\bibitem[{{Villforth} {et~al.}(2010){Villforth}, {Nilsson}, {Heidt}, {Takalo},
  {Pursimo}, {Berdyugin}, {Lindfors}, {Pasanen}, {Winiarski}, {Drozdz},
  {Ogloza}, {Kurpinska-Winiarska}, {Siwak}, {Koziel-Wierzbowska}, {Porowski},
  {Kuzmicz}, {Krzesinski}, {Kundera}, {Wu}, {Zhou}, {Efimov}, {Sadakane},
  {Kamada}, {Ohlert}, {Hentunen}, {Nissinen}, {Dietrich}, {Assef}, {Atlee},
  {Bird}, {Depoy}, {Eastman}, {Peeples}, {Prieto}, {Watson}, {Yee}, {Liakos},
  {Niarchos}, {Gazeas}, {Dogru}, {Donmez}, {Marchev}, {Coggins-Hill},
  {Mattingly}, {Keel}, {Haque}, {Aungwerojwit}, \& {Bergvall}}]{Villforth+10}
{Villforth}, C., {Nilsson}, K., {Heidt}, J., {et~al.} 2010, \mnras, 402, 2087,
  \dodoi{10.1111/j.1365-2966.2009.16133.x}

\end{thebibliography}

\end{document}